\documentclass[12pt]{article}
\usepackage{graphicx} % Required for inserting images
\usepackage[letterpaper, portrait, margin=1in]{geometry}

\usepackage{natbib}
\usepackage[english]{babel}
\usepackage{amsfonts,amsmath,amssymb,amsthm}
\usepackage[capposition=top]{floatrow}
\newfloatcommand{capbtabbox}{table}[][\FBwidth]
\usepackage{cases}
\usepackage{subfig}
\usepackage{threeparttable}
\usepackage{booktabs}
\usepackage{adjustbox}
\usepackage{rotating}
\usepackage{tabularx}
\usepackage{soul}
\newcolumntype{Y}{>{\centering\arraybackslash}X}
\usepackage[table]{xcolor}
\usepackage{enumerate}
\usepackage{float}
\usepackage{comment}
\usepackage{multirow,array}
\usepackage[shortlabels]{enumitem}
\usepackage{rotating}
\usepackage{setspace} 
\usepackage[hidelinks]{hyperref}
\hypersetup{colorlinks,citecolor=blue,linkcolor=blue,urlcolor=black}
\usepackage[hypcap]{caption}
\usepackage[nameinlink]{cleveref}
\usepackage[multiple]{footmisc}
\usepackage{tablefootnote}
\usepackage{adjustbox}
\usepackage{threeparttable}
\usepackage{bbm}
\usepackage{mathtools}
\usepackage{array}
\usepackage[usestackEOL]{stackengine}
\usepackage{graphicx} 
\usepackage{caption}
\usepackage{pifont}
\usepackage{arydshln}
\usepackage{enumitem}
\usepackage{epstopdf}
\usepackage[final]{pdfpages}
\usepackage{eurosym}
\usepackage{tikz}
\usepackage{pgfplots}
\usetikzlibrary{calc}

\newcolumntype{C}[1]{>{\centering\arraybackslash}p{#1}}

\newtheorem{hyp}{Behavioral Hypothesis}

\newtheorem{result}{Result}

\title{Complexity Beyond Incentives: \\The Critical Role of Reporting Language\footnote{\footnotesize We thank Inacio B\'{o}, Dorothea K\"{u}bler, Bozhang Xia and Xi Jin, and participants of Symposium for Theory and Experiment at IES SWUFE, University of Macau, CUHK GBA Conference, and Asia-Pacific Meeting of the ESA 2026 at University of Melbourne for valuable comments and suggestions. We are grateful to Haomin He, Shapeng Jiang, and Wang Jingyi for excellent research assistance. We acknowledge financial support from the National Natural Science Foundation of China (Grant \#W2433170) (Khanna) and Swiss National Science Foundation (Project \#10003543) (Hakimov). The experiment was pre-registered on the AEA registry, number AEARCTR-0013135. The experiment was approved by the LABEX Ethics Board of HEC Lausanne on 08.03.2024, approval no. SAFER.}}

\author{Rustamdjan Hakimov\thanks{University of Lausanne \& WZB Berlin Social Science Center, Internef 536, Quartier de Chamberonne, CH-1015, Lausanne, Switzerland. Email: rustamdjan.hakimov@unil.ch}
\and Manshu Khanna\thanks{Peking University HSBC Business School, Shenzhen 518055, China. Email: manshu@phbs.pku.edu.cn}}

\begin{document}
	\maketitle

\vspace{-0.8cm}

\begin{abstract}
\noindent
Mechanisms specify both allocation rules and message spaces. We study how message spaces affect behavior in a laboratory assignment environment in which objects are bundles of three attributes and preferences are induced by utility formulas. We vary preference complexity and compare full-ranking reports, two attribute-based interfaces, and sequential choice under serial dictatorship. Participants make frequent reporting errors even in a treatment that rewards accurate reporting without any allocation, and errors are more frequent when preferences require trade-offs across attributes. Attribute-based interfaces do not improve accuracy: conditional on what they can express, restricted reports track preferences comparatively well, but representational losses---large for lexicographic reports, small for weighted-attribute reports within our preference domains---offset these gains. Sequential choice yields more accurate assignments and lower efficiency loss and less justified envy; a decomposition attributes roughly one-third of its advantage over full-ranking reporting to the smaller menus that participants face. The results show that the message space affects the performance of strategy-proof assignment mechanisms.
\bigskip 
        
\noindent  \parbox[t][11mm]{.15\linewidth}{\textbf{Keywords:}}
	\parbox[t][11mm]{.84\linewidth}{Market Design, Random Priority,  Multiple Attributes, Complexity, Matching Experiment.} 

    \vspace{-0.2cm}
    
  \noindent  \parbox[t]{.15\linewidth}{\ \ \ \ \ \ \ \  \ \textbf{JEL:}}
	\parbox[t]{.84\linewidth}{C92, D47.}
\end{abstract}

\newpage
\doublespacing

\section{Introduction}
\label{sec:introduction}

Mechanism design is usually cast as the choice of an allocation rule. Yet every mechanism also chooses a \emph{language}: the message space through which participants must communicate their preferences. Standard theory treats the preference report as a free transcription of a known ranking; in practice, the ranking must first be computed and then translated into the mechanism's language. Matching mechanisms allocate scarce resources across many domains—medical residencies, school seats, public housing, organ donations, and limited appointment slots \citep{roth2002economist,abdulkadiroglu2003school,Abdulkadiroglu1999,budish2012multi,roth2004kidney,Hakimov2021}—and their canonical implementations require participants to submit preference rankings over fully specified options.\footnote{Canonical mechanisms—Deferred Acceptance \citep{gale1962college,balinski1999tale}, Random Serial Dictatorship \citep{Abdulkadiroglu1998,Pycia2024}, Top Trading Cycles, and the Cumulative Offer Mechanism \citep{Hatfield2005,hatfield2021}—require agents to submit rankings over fully specified options. Preferences in standard models are typically assumed to be known; a few studies relax this assumption by allowing costly preference learning \citep{Chen2021,Chen2022,Hakimov2023a}, while others allow for indecisiveness \citep{Caspari2024}.} When options are defined by multiple attributes, i.e., levels of various attributes uniquely define an object (e.g., university $\times$ major $\times$ tuition), this seemingly simple task becomes cognitively demanding, taxing attention, memory, and computation \citep{rees2018experimental,Milgrom2009,Milgrom2011}. The burden is evident in practice: Chinese applicants rank hundreds of institution–major bundles \citep{huchinesecolleges}; U.S. Military Academy cadets evaluate branch–service combinations \citep{greenberg2024redesigning,sonmez2013bidding}; and bidders in combinatorial auctions assess packages with complementarities \citep{Parkes2005,Kwasnica2005,sandholm2005}. The question of how the choice of reporting language shapes what mechanisms actually elicit has received surprisingly little attention.

We address two central questions. First, is ranking multi-attribute options inherently difficult, and does this difficulty vary with preference structure? Second, do simplified reporting interfaces—widely adopted in practice—actually help participants communicate their preferences more accurately? 

These questions matter because many real-world systems restrict how preferences can be expressed, presumably to reduce cognitive burden. China's college admissions system, for example, uses a structured rank-order format that requires ranking universities first, then within each university, ranking the majors offered by that institution. This nesting of majors under institutions effectively enforces lexicographic preferences \citep{huchinesecolleges}.\footnote{As of January 2025, 23 of 31 provinces maintain this format. Appendix B shows sample forms from Fujian and Shanghai.}
 At the U.S. Military Academy (USMA), \citet{sonmez2013matching} proposed replacing the USMA-2006 mechanism with a more expressive alternative, but the Academy retained USMA-2006 to avoid excessive reporting burden \citep{greenberg2024redesigning}.\footnote{\cite{greenberg2024redesigning} write ``While this \citep{sonmez2013matching} proposal had desirable theoretical properties, it required a more complex strategy space in which cadets have to rank branches and contractual terms (also referred to as prices) jointly. Under the USMA-2006 mechanism, cadets only rank branches and separately indicate their willingness to BRADSO for any branch. The Army considered the existing strategy space manageable compared to a more complex alternative and kept the USMA-2006 mechanism in the intervening years." } Other simplification strategies include bidding with artificial budgets \citep{Budish2011,Budish2022}, constrained choice sets \citep{haeringer2009constrained,calsamiglia2010constrained,Huang2025}, sequential menu choice \citep{Bo2024,mackenzie2022menu}, and AI-guided elicitation \citep{soumalias2025llm}. While these approaches reduce cognitive load, they may restrict expressiveness and lower efficiency.
 
We study these questions in laboratory experiments with student subjects. Our experimental design mimics a stylized college admissions setting where 27 participants are assigned to 27 program seats. Each seat is defined by three attributes—university, field of study, and tuition.\footnote{Tuition is a relevant attribute in systems where the same seat may be offered with or without tuition, such as in Israel and Hungary \citep{hassidim2021limits,shorrer2023}.} Each attribute has three levels, yielding $3 \times 3 \times 3 = 27$ distinct programs (bundles). Students know the distribution from which exam scores are drawn, and their exam scores, and allocations are determined by Serial Dictatorship in most treatments. We independently vary (i) the complexity of preferences over attributes and (ii) the interface used to report preferences.

Our design has three innovations. First, unlike most matching experiments, we do not assign ordinal preferences directly.\footnote{Exceptions include \citet{Budish2022}, where participants express home grown preferences, and the preference information is elicited through a series of binary choices;  \citet{kloosterman2023rankings} who use real objects; and \citet{guillen2018effectiveness} run a field experiment with real preferences of students over topics.} Instead, each participant is given a utility function that maps attribute profiles into utility scores. These scores define ordinal rankings over programs, and payoffs depend on the rank of the assigned program. This allows us to manipulate complexity of preferences over attributes exogenously\footnote{It also reduces experimenter demand effect, since participants cannot simply copy pre-filled preference tables to be truthful.}—ranging from lexicographic (LEX), to additive separable (SEP), to non-separable with attribute interactions (COMP).\footnote{In LEX, university is the most important attribute, dominating field of study, which in turn dominates tuition; that is, any program in the top-ranked university is preferred to any program in the second-ranked university, and so on. In SEP, no attribute dominates: each attribute enters with a weight, and the program score increases linearly in the values of the three attributes. In COMP, in addition to attribute weights, there is an interaction term between university prestige and tuition---which can be positive or negative---so the relationship between program scores and attributes is no longer linear.}
 This treatment dimension captures cognitive challenges in forming and articulating preferences.

Second, to separate cognitive barriers to truthful reporting from incentive-driven misreporting in a strategy-proof mechanism,\footnote{In strategy-proof mechanism, there are no incentives to misreport, but empirical evidence  show that this is common \citep{hassidim2021limits}.} we include a baseline treatment (ACCURACY). Here, participants are rewarded for reporting true rankings of 27 bundles, but no allocation takes place. This non-allocative full-ranking benchmark provides a measure of reporting frictions, abstracting from strategic incentives of the mechanism.

Third, we vary reporting interfaces, i.e., how preferences are communicated to the Serial Dictatorship mechanism. In the direct interface (SD-DIRECT), participants submit a full ranking over all 27 bundles. The lexicographic interface (SD-LEX) asks participants to rank each attribute separately\footnote{Under SD-LEX, participants rank the three universities, the three fields, and the three tuition levels separately, instead of submitting a single 27-item ranking required in SD-DIRECT.}
 and constructs full bundle rankings for Serial Dictatorship assuming lexicographic preferences (university $\succ$ field $\succ$ tuition). This mirrors systems like China’s, where the  lexicographic structure over attributes is exogenously imposed: it fits lexicographic preferences exactly but distorts rankings when preferences are more complex. A third interface (SD-WEIGHT) also asks participants to rank attributes separately, but allows weights across attributes.\footnote{Participants submit three rankings of three items, as in SD-LEX, but additionally specify a weight for each attribute.}  Expressiveness differs across interfaces.\footnote{In this context, \textit{expressiveness} refers to the ability of a reporting interface to accurately represent all possible preference orderings over the available options. A \textit{fully expressive} interface allows participants to communicate any ranking of the 27 programs. An interface has limited expressiveness when certain preference orderings cannot be accurately represented through that format.} 

Under LEX preferences, all three interfaces (SD-LEX, SD-WEIGHT, and SD-DIRECT) are fully expressive; under COMP and SEP, neither SD-LEX nor SD-WEIGHT is fully expressive.\footnote{Initially, we hypothesized that SD-WEIGHT would be fully expressive under SEP preferences, since SEP involves additive utilities across attributes and SD-WEIGHT allows participants to assign importance weights to different attributes. However, SD-WEIGHT cannot represent all possible SEP preferences because the attribute values in our setting are not uniformly spaced. The SD-WEIGHT interface implicitly assumes uniform spacing between ranked items within each attribute, but when actual attribute values have non-uniform gaps, the weighted sum of ranks cannot capture all possible additive utility functions that SEP preferences allow. Section~\ref{sec:expressiveness} quantifies the expressiveness of each interface across preference domains: SD-LEX and SD-WEIGHT are both fully expressive on LEX, but neither can express all SEP or COMP preferences, and the magnitude of the expressiveness loss differs sharply between the two interfaces.} Thus, while SD-LEX and SD-WEIGHT reduce the burden of submitting a full 27-item ranking, they necessarily introduce misspecification under more complex preference structures, where attribute trade-offs cannot be represented by lexicographic or weighted-attribute aggregation.\footnote{Thus, practical relevance of SD-LEX and SD-WEIGHT depends not only on how they match underlying preferences, but also on how high are the behavioral benefits of simplified reporting relative to SD-DIRECT.}

Additionally, we study a mechanism that changes both interface and strategy space relative to SD-DIRECT: sequential serial dictatorship (SD-CHOICE). Here, each participant chooses her most preferred available option when her turn arrives in the priority order. SD-CHOICE requires no full ranking and is \emph{obviously strategy-proof} \citep{Li2017,Pycia2023}. It thus serves dual roles—as a simpler interface and as a mechanism with stronger incentive properties. Comparing SD-CHOICE to SD-DIRECT measures the combined gains from obvious strategy-proofness and the simpler interface---a gap we decompose below---while comparing SD-CHOICE to ACCURACY asks whether its simplified interface can outperform even a non-allocative full-ranking benchmark that removes strategic incentives altogether.

We conducted the experiment with 810 university students, randomly assigned across treatments. Preference complexity was varied within subjects, while reporting interfaces and mechanisms were varied between subjects. Each session included 12 repeated rounds covering the different complexities in random order. Depending on the treatment, participants’ final payoffs were determined either by the outcomes of the resulting serial-dictatorship allocations or by the accuracy of their reported preferences (ACCURACY).

Our results yield five main findings. First, in ACCURACY, with no strategic incentives, participants frequently misreport their preferences, and misreporting is substantially more frequent in the trade-off domains (SEP and COMP) than under lexicographic preferences.\footnote{We use two measures of accuracy. One is choice accuracy, a dummy equal to one if the participant chooses (or ranks first) the best program among those available when it is her turn to choose or to be allocated a seat. The second is the normalized Kendall--tau distance between the submitted and true rankings of the 27 programs.}
 Reporting accuracy is significantly higher in ACCURACY-LEX than in ACCURACY-SEP or ACCURACY-COMP. This highlights that ordinal reporting over many alternatives is cognitively demanding, especially when attribute-based preferences are complex. The negative effect of complexity on truth-telling is replicated across all other treatments (SD-DIRECT, SD-LEX, SD-WEIGHT, SD-CHOICE).

Second, in the direct Serial Dictatorship treatments (SD-DIRECT-LEX, SD-DIRECT-SEP, SD-DIRECT-COMP), misreporting is widespread and consistent with incentive-related and framing-related reporting errors. Error rates are significantly higher than in the corresponding ACCURACY treatments, which remove mechanism incentives. This complements earlier evidence on misreporting in the serial dictatorship mechanism \citep{Li2017,Bo2024}, showing that it is not solely noise or a consequence of the length of the rank-order list.

Third, simplifications in the reporting language (SD-LEX and SD-WEIGHT) do not improve accuracy of reporting. These reporting interfaces require three separate rankings (over universities, fields, and tuition levels) instead of a full ranking of 27 bundles. Yet overall the accuracy of reporting is significantly lower than in SD-DIRECT. Even under lexicographic preferences, SD-LEX does not outperform SD-DIRECT (accuracy rates are statistically indistinguishable between SD-LEX-LEX and SD-DIRECT-LEX, and significantly lower in SD-WEIGHT-LEX). Simplification thus fails to improve accuracy, even when the reporting interface matches the underlying preference structure. Our expressiveness accounting sharpens this finding: conditional on what each interface can express, restricted reports track the best expressible ranking comparatively well; what offsets these gains are representational restrictions---substantial for SD-LEX, modest for SD-WEIGHT---and neither interface improves assignment accuracy.

Fourth, the sequential version of Serial Dictatorship (SD-CHOICE) has the highest accuracy and the lowest efficiency loss and justified envy in our experiment. SD-CHOICE exceeds ACCURACY overall, with the clearest difference in SEP. This result suggests that SD-CHOICE improves outcomes not only because of stronger incentive properties, but also because it changes the reporting interface. A reduced-form decomposition of the SD-CHOICE–SD-DIRECT gap attributes roughly one-third to $45\%$ to a priority-position/menu component---participants choosing from smaller menus---and the remainder to a residual SD-CHOICE component that includes obvious strategy-proofness and the simpler one-at-a-time decision frame, with the menu component stable across preference domains. Its success therefore reflects both incentive simplification and a quantitatively meaningful interface channel.

Finally, reporting accuracy translates into economically meaningful outcomes. Treatments with higher truthfulness generate fairer allocations, as reflected in lower justified envy; welfare gains are concentrated in the comparisons of ACCURACY and SD-CHOICE with the direct interfaces, whose efficiency losses are similar to one another. Misreporting is thus not confined to irrelevant parts of preference lists, but arises in consequential parts that directly affect allocations.

Overall, our findings highlight the challenges of accurately reporting multi-attribute preferences and show how complexity interacts with cognitive and strategic factors to generate misreporting. Sequential Serial Dictatorship (SD-CHOICE) performs best in our experiment, combining simple incentives with a simple interface. When sequential implementation is not feasible, our evidence favors the full-ranking interface over attribute-based formats, which fail to improve accuracy even when aligned with the underlying preference domain.

\textbf{Literature}. This paper relates to four strands of research. First, work on the costs of eliciting multi-attribute preferences shows that rich reporting is hard both technologically and cognitively. \citet{Compte2004} formalize limits to communication and elicitation when preferences are high-dimensional. In combinatorial environments, \citet{Parkes2005} shows how package structure and complementarities strain elicitation and computation. From a behavioral angle, \citet{Milgrom2009,Milgrom2011} argues that multi-attribute trade-offs tax attention and working memory, predicting systematic error and heuristic reporting. In large-scale course allocation, \citet{Budish2011} introduces an approximate competitive equilibrium mechanism (A-CEEI) that requires bidding for courses with fake money instead of ranking course bundles, and \citet{Budish2022} document that participants can communicate preferences accurately enough to realize the efficiency and fairness benefits of A-CEEI. But preference-reporting mistakes are common and meaningfully harm mechanism performance. \citet{sonmez2010course} provide further evidence that course-assignment environments induce nontrivial reporting frictions. We build on this strand by separating the complexity of underlying preferences from the complexity of the reporting task and by benchmarking reports against a known ground truth.

Second, a growing literature documents misreporting under strategy-proof mechanisms and analyzes its sources both in the lab and in the field \citep{Chen2006,Hakimov2021a,rees2023behavioral}. Using residency data, \citet{ReesJones2018b} shows sizable deviations from truth-telling in dominant-strategy environments; \citet{hassidim2021limits,chen2019self,Shorrer2024} demonstrate that even well-incentivized centralized procedures exhibit systematic non-truthful play. Another line of papers tries to rationalize reasons behind the deviations, such as reference dependence, rank utility, disappointment aversion, and others \citep{dreyfuss2022deferred,chen2024does,meisner2023loss,Meisner2023,kloosterman2023rankings}. \citet{gonczarowski2023strategyproofness,gonczarowski2024describing} demonstrate that the way mechanisms are explained—particularly whether strategy-proofness is made salient—can significantly affect behavior. Similarly, \citet{katuvsvcak2024strategy} show that simplifying the explanation of strategy-proof mechanisms increases truthful reporting. Finally, \citet{guillen2021strategy} caution that lab estimates of truth-telling may be inflated due to experimenter demand effects, where participants conform to perceived expectations. Our contribution is to show that these deviations are not mere noise: holding the choice set fixed, misreporting rises with preference complexity; holding complexity fixed, misreporting is higher under direct serial dictatorship than in a non-allocative accuracy benchmark, consistent with incentive-related and framing-related errors rather than list length alone.

Third, studies compare sequential (iterative) to direct implementations and quantify the role of menu size and feedback. In the lab, \citet{Klijn2019,bo2020iterative} find that iterative deferred acceptance improves stability and welfare relative to direct submissions; \citet{mackenzie2022menu,Bo2024} show theoretically and in experiments that staging choices raises truthful play and assignment quality; and \citet{Haeringer2021} analyze iterative designs that simplify participation. \citet{Stephenson2022} shows that real-time assignment feedback during reporting increases equilibrium behavior in school choice. \citet{Dur2021} analyze sequential Boston mechanisms, finding similar improvements in participant outcomes. In higher-education admissions, \citet{Hakimov2023a} show that sequential serial dictatorship facilitates information acquisition about preferences. Field evidence from \citet{Grenet2022} shows the importance of a dynamic multi-offer mechanism for preference formation in Germany. \citet{Gong2024} provide theory and experiment for a dynamic admissions mechanism that achieves efficient and stable outcomes under mild conditions. On incentives, \citet{Li2017} establishes that a sequential version of Random Priority is obviously strategy-proof, and \citet{Pycia2023} develop a broader simplicity framework clarifying why shrinking menus reduce incentive misperception. We contribute to this literature by documenting the superior performance of sequential serial dictatorship, and by showing that part of its advantage is driven by a simpler interface—smaller menus—rather than by stronger incentives alone.

Finally, platforms often restrict the message space to ease reporting, trading expressiveness for lower burden. One example is constrained lists, which are prevalent in practice worldwide, despite clear theoretical and experimental evidence that they harm incentives \citep{haeringer2009constrained,calsamiglia2010constrained}. In practice, lexicographic formats are used to structure reporting: \citet{huchinesecolleges} document institution–major submission in China, while \citet{greenberg2024redesigning} describe a minimalist redesign for Army branching that limits what must be reported to meet policy goals. Relatedly, \citet{aygun2026affirmative} show, in the context of Indian affirmative action with hierarchical reservations, that enriching the preference domain to allow expression over institution–vertical-category pairs substantially changes the set of implementable allocations—underscoring that the choice of preference domain, like the choice of reporting language we study, is a consequential design lever. Alternative simplifications include budget-based inputs for course allocation \citep{Budish2011,Budish2022}. Our experiments quantify the trade-off: attribute-based simplifications (lexicographic and weighted attributes) do not improve accuracy even when they match the preference domain.

The remainder of the paper proceeds as follows. Section~\ref{sec:setup} details our experimental design and procedures. Section~\ref{sec:results} presents the key findings across reporting languages and preference structures. Section~\ref{sec:conclusion} concludes and offers implications for the design of multi-attribute matching mechanisms.

\section{Experiment Design and Behavioral Hypotheses}
\label{sec:setup}

\subsection{Environment}

\subsubsection{Admissions Scenario}
In each round, participants acted as applicants in a simulated college admissions market. Each market consisted of 27 students and 27 programs, with one seat per program. Programs were defined by three attributes: university prestige, field of study, and tuition. Attribute values were fixed as follows: prestige points were 500 (University A), 200 (University B), and 600 (University C); tuition points were 0 (no tuition), 250 (half tuition), and 500 (full tuition); field-of-study points were $\{300,500,700\}$, with Economics, Finance, and Law randomly assigned to these values each market.

Each participant received a \emph{preference-score formula} (utility function) mapping attribute values into scores for every program. Higher scores implied higher positions in the participant’s induced ordinal ranking. An on-screen calculator allowed participants to compute scores for any program. Our design replaces per-seat monetary payoffs with these formulas: scores induce the participant’s true ordinal ranking, and payoffs depend only on the rank of the assigned program, not on cardinal scores.\footnote{We avoid the usual assignment of monetary payoffs to individual programs: with per-program payoffs, reporting even a long ranking becomes trivial and leaves no room to vary the complexity of preferences. These reasons are similar to discussion of \citet{Budish2022} on the unsuitability of inducing preferences in a format immediately reportable to the mechanism. However, we deliberately adopt the approach that is criticized by \citet{Budish2022}: one “language” assigns incentives (utility formulas), and participants must translate it into the reporting language of the mechanism. While we agree with the general criticism, this serves our purpose: it allows us to separate the complexity of \emph{inducing} preferences from the complexity of \emph{reporting} them, while holding preferences fixed.} This ensures comparable incentives across treatments and rounds and allows us to vary preference complexity while abstracting from strategic motives.

In the SD-DIRECT, SD-WEIGHT, and SD-LEX treatments, priorities were determined by exam marks (uniformly drawn without replacement from $\{1,\ldots,100\}$ to avoid ties). Participants knew their own mark but not others’. In SD-CHOICE, priorities were revealed sequentially as participants were called to act in order. In ACCURACY, no priority order was used, as there was no assignment. Each session consisted of 12 rounds, with new preference formulas and marks each round.

\subsubsection{Mechanisms}
In SD-DIRECT, SD-WEIGHT, and SD-LEX, programs were allocated via standard serial dictatorship. The mechanism uses participants' submitted rank-order lists of programs as input. The participant with the highest mark received her top-ranked program; the participant with the second-highest mark received her top-ranked program among those still available, and so on.

In SD-CHOICE, participants did not submit rankings. The student with the highest priority selected her most-preferred program from all 27 programs; the student with the second-highest priority then selected from the remaining programs, and so on.\footnote{To keep 12 rounds while minimizing time, we ran them in parallel. Conceptually, there were 12 independent admissions processes. Each of 27 “turns” advanced all 12 processes by one pick: we randomly selected 12 participants (one per process) to choose in the current turn while the other 15 waited. After 27 turns, all 12 processes were complete and each participant had made exactly one choice in each process.}

In ACCURACY, no allocation took place. Participants were rewarded purely for reporting rankings that matched their true preferences.\footnote{ACCURACY is not a frictionless benchmark: participants must still compute preference scores, construct a ranking of 27 programs, and understand the Kendall-distance payoff rule. We therefore interpret it as measuring reporting frictions in the absence of strategic incentives, not as a zero-error benchmark.} Accuracy was measured using the normalized Kendall distance, the fraction of discordant pairs between the reported and true rankings:
\[
\text{Kendall distance} \;=\; \frac{\text{number of discordant pairs}}{\text{total number of pairs}}.
\]

 The measure ranges from $0$ (identical rankings) to $1$ (complete reversal).\footnote{Example: Ranking 1 is $A,B,C,D$; Ranking 2 is $B,A,D,C$. The discordant pairs are $(A,B)$ and $(C,D)$, so there are $2$ discordances out of $\binom{4}{2}=6$ possible pairs, giving a normalized Kendall distance of $2/6=0.333$.}

\subsubsection{Payoffs}
In all treatments, one of the 12 rounds was randomly selected for payment. In all treatments except ACCURACY, participants received 160 Chinese yuan (CNY) if assigned their truly most-preferred program (based on their induced preference formula), CNY~155 for their true second-ranked program, and so on, decreasing by CNY~5 per rank to CNY~30 for the least-preferred program.\footnote{Participants received a payoff table in the instructions (Appendix~C).} In ACCURACY, the payoff was
\[
\text{Payoff} \;=\; 160 \times \bigl(1 - \text{Kendall distance}\bigr),
\]
so closer agreement with the true ranking yielded higher earnings.

\subsection{Treatment Variations}
We implement a $3\times 5$ design combining three levels of preference complexity with five reporting interfaces/mechanisms. Preference complexity varies \emph{within} subjects across rounds; the reporting interface/mechanism varies \emph{between} subjects. We now describe both dimensions.

\subsubsection{Complexity of Preferences}\label{sec:comppref}
Participants faced three preference domains induced by preference-score formulas. Let $U\in\{200,500,600\}$ denote university prestige points, $F\in\{300,500,700\}$ field-of-study points (major-to-points mapping randomized each market), and $T\in\{0,250,500\}$ tuition points (higher $T$ means higher tuition). For each subject and round, coefficients are drawn independently and uniformly from the indicated intervals.

\begin{enumerate}
\item \textbf{Lexicographic (LEX).} Scores are
\[
s = a\,U + b\,F - c\,T,\quad a\in[90,110],\; b\in[9,11],\; c\in[0.9,1.1].
\]
The order-of-magnitude separation of coefficients enforces the lexicographic priority $U \succ F \succ -T$ on the feasible attribute ranges, so differences in $U$ dominate any differences in $F$ or $T$, and differences in $F$ dominate $T$.

\item \textbf{Additively separable (SEP).} Scores are
\[
s = a\,U + b\,F - c\,T,\quad a,b,c\in[30,40],
\]
so attributes carry comparable weights and trade-offs must be computed across all three dimensions.

\item \textbf{Non-separable with attribute interactions (COMP).} Scores are
\[
s = a\,U + b\,F - c\,T + d\,U\cdot T,\quad a,b,c\in[30,40],\; d\in[-5,5],
\]
introducing an interaction between prestige and tuition. Positive $d$ weakens the tuition penalty at high-prestige universities (willingness to pay for prestige); negative $d$ strengthens it. The interaction applies to the attribute values exactly as shown to participants; because $U$ and $T$ are in the hundreds, the interaction term can be large relative to the linear terms, so that for some draws it materially changes the ranking and can reverse the tuition ordering across universities.
\end{enumerate}

The 12 rounds are organized into four blocks of three rounds (except in SD-CHOICE due to the parallel implementation described above). In the first three rounds, participants face all three preference types (LEX, SEP, COMP), one per round in random order. In the next three rounds, they again face all three preference types in a different random order, and this pattern continues for the subsequent two blocks of three rounds. The block design allows for a balance of experience across preference domains.

\subsubsection{Preference Reporting Languages}
\begin{enumerate}
\item \textbf{SD-DIRECT.} After observing their exam mark, participants submit a full ranking of the 27 programs.

\item \textbf{SD-WEIGHT.} Participants rank each attribute (university, field of study, tuition) separately and report integer attribute weights in $[1,100]$. A composite “points” index is computed and minimized to produce the implied ranking of 27 programs.\footnote{Displayed to participants: 
\[
\text{Points} = \text{rank}(U)\times w_U + \text{rank}(F)\times w_F + \text{rank}(T)\times w_T.
\]
Lower points imply a higher bundle rank. Example: if $U{=}A$, $F{=}$Economics, $T{=}$No tuition are ranked first, and Finance is ranked second for $F$, with $(w_U,w_F,w_T)=(50,20,70)$, then
$[A,\text{Economics},\text{No tuition}]$ has $1\times50+1\times20+1\times70=140$, whereas $[A,\text{Finance},\text{No tuition}]$ has $1\times50+2\times20+1\times70=160$.}

\item \textbf{SD-LEX.} Participants rank each attribute separately; the ranking of 27 programs is constructed by imposing a lexicographic structure with university $\succ$ field $\succ$ tuition.\footnote{Displayed to participants:
\[
\text{Points} = 100\times\text{rank}(U) + 10\times\text{rank}(F) + 1\times\text{rank}(T).
\]
Lower points imply a higher bundle rank. Example: if $U{=}A$, $F{=}$Economics, $T{=}$No tuition are ranked first, and Finance is ranked second for $F$, then
$[A,\text{Economics},\text{No tuition}]$ has $1\times100+1\times10+1\times1=111$, while
$[A,\text{Finance},\text{No tuition}]$ has $1\times100+2\times10+1\times1=121$.}

\item \textbf{SD-CHOICE.} Participants act in order of priority. When called, a participant observes the remaining programs and selects one.\footnote{Participants were allowed to submit an empty choice; doing so yielded zero payoff for that round.}

\item \textbf{ACCURACY.} No allocation occurs. Participants submit a ranking of the 27 programs and are paid based on the (normalized) Kendall distance between their submitted and true rankings.
\end{enumerate}

\subsection{Procedures}

We ran the experiments at Wuhan University’s Research Center for Behavioral Science during 2024–2025 with 810 Wuhan University students. Table~\ref{tab:treatments} summarizes the sessions by treatment. For each treatment, we conducted six sessions with 27 participants each (one independent matching group per session), yielding 162 participants per treatment. Participants were randomly assigned to one treatment and took part in only that treatment. Payments were made privately by transfer at the end of the session.

\setlength{\tabcolsep}{3pt}
\renewcommand{\arraystretch}{1.2}
\begin{table}[htbp]
\centering \footnotesize
\caption{Experimental Design} 
\label{tab:treatments}
\begin{tabular}{lccll}
\toprule
Treatment & Participants & Sessions & Reporting Interface & Allocation Mechanism \\
\midrule
ACCURACY   & 162 & 6 & Full ranking & None; incentivized truthful reporting \\ 
SD-DIRECT  & 162 & 6 & Full ranking & Serial Dictatorship  \\ 
SD-WEIGHT  & 162 & 6 & Attribute rankings $+$ weights & Serial Dictatorship  \\
SD-LEX     & 162 & 6 & Attribute rankings & Serial Dictatorship  \\ 
SD-CHOICE  & 162 & 6 & Pick when called & Sequential Serial Dictatorship \\
\bottomrule
\end{tabular} 
\end{table}

At the beginning of each session, the 27 participants were given the written experimental
instructions, allowing them to follow along as the experimenters read the instructions aloud. The instructions described the environment, allocation procedures, and payoffs. A countdown timer allowed three minutes per choice in SD-CHOICE and eight minutes per reporting round in the other treatments.
Sessions lasted on average 112 minutes. Average earnings were CNY~113.65 with no show-up fee, above typical payments in China.

\paragraph{Preregistration.} The experiment was preregistered at the AEA RCT Registry (AEARCTR-0013135). The treatment structure, the main outcome measures, and the primary treatment comparisons were specified before data collection. The expressiveness calculations in Section~\ref{sec:expressiveness}, the interface-penalty decomposition, and the reduced-form decomposition of the SD-CHOICE advantage in Section~\ref{sec:sd-choice-decomp} were developed after data collection and should be interpreted as mechanism analyses of the experimental results.

\subsection{Interface Expressiveness}
\label{sec:expressiveness}

Before turning to behavior, we quantify how restrictive each reporting
interface is. ACCURACY and SD-DIRECT allow participants to express any
of the $27! \approx 1.09 \times 10^{28}$ possible orderings of programs,
and SD-CHOICE imposes no restriction either, since each choice can
condition on the realized menu.
SD-LEX, by contrast, can produce only the $6 \times 6 \times 6 = 216$ 
rankings induced by combining permutations of the three universities, 
three fields, and three tuition levels with the lexicographic aggregator 
$U \succ F \succ T$. SD-WEIGHT, with integer weights in $[1,100]$ applied
to attribute ranks, can produce substantially
more rankings than SD-LEX.

To translate this into a behavioral benchmark, we simulate $10{,}000$ 
preference draws from each domain (LEX, SEP, COMP) using the score 
formulas in Section~\ref{sec:comppref}, and for each draw we compute
the minimum normalized Kendall distance between the induced true ranking
and the closest ranking the interface can express. Table~\ref{tab:expressiveness}
reports the results; Appendix~\ref{app:expressiveness} details the computation and a robustness check using the realized preference profiles of the experiment.

\begin{table}[h]
\centering
\caption{Interface Expressiveness}
\label{tab:expressiveness}
\begin{tabular}{lrrrr}
\toprule
Interface & Distinct rankings & LEX & SEP & COMP \\
\midrule
SD-DIRECT  & $27! \approx 1.09 \times 10^{28}$ & 0 & 0 & 0 \\
SD-WEIGHT  & 58{,}176 & 0 & 0.02 & 0.09 \\
SD-LEX     & 216    & 0 & 0.23 & 0.29 \\
\bottomrule
\end{tabular}

\vspace{0.5em}
\begin{minipage}{0.9\textwidth}
\footnotesize \textit{Notes:} The second column reports the number of distinct full rankings of 27 programs that each interface can express. Columns 3--5 report the average minimum normalized Kendall distance between true rankings drawn from each preference domain and the closest ranking achievable under the interface, averaged across $10{,}000$ simulated draws. A value of $0$ means the interface is fully expressive for that domain.
\end{minipage}
\end{table}

Three points stand out. First, both SD-LEX and SD-WEIGHT are fully expressive on the lexicographic domain (zero expressiveness cost). Second, SD-WEIGHT's expressiveness cost is small ($\leq 0.09$), confirming that---despite its simple form---it is close to fully expressive on these domains. Third, SD-LEX imposes a sizeable mechanical distortion under SEP ($0.23$) and COMP ($0.29$): even a participant who optimizes perfectly within the interface cannot reduce the Kendall distance below this floor. The behavioral implications of these expressiveness floors are analyzed in Section~\ref{sec:interfaces}.

\subsection{Conceptual Framework and Predictions}\label{sec:hypotheses}

We test how preference complexity and the reporting interface/mechanism affect reporting accuracy and allocation outcomes.

To simplify exposition, we first introduce notation.  

Let $\mathcal{T}=\{\text{ACCURACY},\text{SD-DIRECT},\text{SD-LEX},\text{SD-WEIGHT},\text{SD-CHOICE}\}$ be the between-subjects treatments and $\mathcal{D}=\{\text{LEX},\text{SEP},\text{COMP}\}$ the preference domains. For $t\in\mathcal{T}$ and $d\in\mathcal{D}$, write $\mathrm{Acc}(t,d)$ for expected reporting accuracy.

First, we hypothesize that the complexity of preferences hurts outcomes. Prior work shows that higher task complexity raises errors \citep{kahneman2003maps,gilovich2002heuristics}; non-separabilities increase comparison difficulty \citep{Milgrom2009,Milgrom2011}; and preference reporting can be costly \citep{Parkes2005,Budish2022,sonmez2010course}.

The three preference domains differ in the computations required to rank bundles. In LEX, the coefficient separation implies a fixed hierarchy of attributes, so most comparisons can be resolved without computing close trade-offs. In SEP, no attribute dominates, and participants must compare additive trade-offs across university, field, and tuition. In COMP, the value of tuition also depends on university prestige, so the relevant trade-off varies across the choice set. We therefore expect LEX to generate more accurate reports than the two trade-off domains. The comparison between SEP and COMP is less sharp ex ante: COMP is richer, but participants may respond by spending more time or relying more heavily on the on-screen calculator.

\begin{hyp}[Preference complexity reduces reporting accuracy]\label{hyp:complexity}
Fix any between-subjects treatment $t\in\mathcal{T}$. Reporting accuracy is higher in LEX than in the trade-off domains:
\[
\mathrm{Acc}(t,\text{LEX}) \;>\; \mathrm{Acc}(t,\text{SEP})
\quad\text{and}\quad
\mathrm{Acc}(t,\text{LEX}) \;>\; \mathrm{Acc}(t,\text{COMP}).
\]
The difference between SEP and COMP is ambiguous ex ante and is treated as exploratory.
\end{hyp}

Second, we hypothesize that rates of misreporting in the direct serial dictatorship mechanism are higher than in the ACCURACY treatment. Such a gap would indicate that previously documented misreporting in serial dictatorship is nonrandom and consistent with incentive-related misreporting, since ACCURACY removes allocation incentives---although, because the two treatments also differ in their payoff rules, the gap is an incentive-and-framing contrast rather than a pure measure of strategic misreporting.

\begin{hyp}[Additional reporting errors under direct serial dictatorship]\label{hyp:misperception}
For each preference domain $d\in\mathcal{D}$,
\[
\mathrm{Acc}(\text{ACCURACY},d) \;>\; \mathrm{Acc}(\text{SD-DIRECT},d).
\]
\end{hyp}

Third, we consider the attribute-based reporting interfaces. Relative to SD-DIRECT, their net effect reflects the balance of three forces: a reduced input burden (three rankings of three items instead of one ranking of 27 programs), a loss of expressiveness (absent on LEX, and far larger for SD-LEX than for SD-WEIGHT on SEP and COMP; Table~\ref{tab:expressiveness}), and the cost of using the interface itself---translating preferences into attribute rankings and weights \citep{Milgrom2009,Milgrom2011}. The first force favors the simplified interfaces; the other two work against them. Because only the expressiveness cost varies sharply across domains and interfaces, the balance delivers directional predictions across domains, while on the LEX domain itself---where all interfaces are fully expressive---the sign of the comparison is ambiguous ex ante: it depends on whether the reduced input burden outweighs interface-use costs.

\begin{hyp}[Attribute-based interfaces: burden reduction versus expressiveness and use costs]\label{hyp:simplifications}
The performance of the simplified interfaces relative to SD-DIRECT deteriorates across domains in line with their expressiveness loss. For SD-LEX, whose loss is large on SEP and COMP,
\[
\begin{aligned}
&\mathrm{Acc}(\text{SD-LEX}, \text{LEX}) - \mathrm{Acc}(\text{SD-DIRECT}, \text{LEX}) \\
&\qquad >\; \mathrm{Acc}(\text{SD-LEX}, d) - \mathrm{Acc}(\text{SD-DIRECT}, d), \qquad d \in \{\text{SEP},\text{COMP}\},
\end{aligned}
\]
and $\mathrm{Acc}(\text{SD-LEX},d) \le \mathrm{Acc}(\text{SD-WEIGHT},d)$ for $d\in\{\text{SEP},\text{COMP}\}$. For SD-WEIGHT, whose loss is small, relative performance should vary little across domains.
\end{hyp}

Fourth, SD-CHOICE simplifies both incentives—making the mechanism obviously strategy-proof—and the reporting interface by narrowing the active choice set \citep{Li2017,Pycia2023,chernev2015choice}. We therefore expect improvements over direct serial dictatorship. The comparison with ACCURACY is ambiguous ex ante: ACCURACY removes allocation incentives but still requires constructing a full ranking, whereas SD-CHOICE preserves allocation consequences but replaces full-list reporting with a sequence of one-item choices. If the burden of full-list construction is large, SD-CHOICE may match or exceed ACCURACY, especially in more complex domains. Section~\ref{sec:sd-choice-decomp} decomposes the SD-CHOICE advantage into a menu-size component and a residual.

\begin{hyp}[Sequential choice improves accuracy over direct reporting]\label{hyp:sequential}
For each domain $d\in\mathcal{D}$,
\[
\mathrm{Acc}(\text{SD-CHOICE},d) \;>\; \mathrm{Acc}(\text{SD-DIRECT},d).
\]
The comparison with ACCURACY is ambiguous ex ante; if the burden of full-list construction is large, SD-CHOICE may match or exceed ACCURACY, especially in more complex domains.
\end{hyp}

Finally, while all hypotheses are formulated about accuracy, we also analyze efficiency and justified envy of the resulting allocations in Section \ref{sec:efficiencyenvy}. We expect deviations from accurate reporting to occur in parts of the preference ranking that matter for allocation given priorities. Accordingly, treatment differences in reporting accuracy should lead to analogous differences in efficiency and justified envy.

\section{Results}
\label{sec:results}

%----------------------------------------------------------------------
% OVERVIEW
%----------------------------------------------------------------------

Unless stated otherwise, statistical significance is assessed at the 5\% level. The regressions include random effects at the matching-group and participant levels, with standard errors clustered at the matching-group level; for binary outcomes, we report average marginal effects from logistic regressions. For pairwise treatment comparisons, we report $p$-values on treatment coefficients from regressions run on the relevant two-treatment subsample, with standard errors clustered at the matching-group level. The full set of pairwise comparisons, overall and by preference domain, is reported in Appendix Tables~\ref{tab:pairwise}--\ref{tab:pairwise_r4}, which display log-odds coefficients. Because treatment is assigned at the matching-group level and each treatment has six independent groups, we additionally report exact randomization-inference $p$-values for the headline pairwise comparisons, permuting treatment labels across the twelve matching groups of each comparison (Appendix Table~\ref{tab:ri}). Round-by-round treatment profiles of all four outcome measures are shown in Appendix Figures~\ref{fig:rounds-acc}--\ref{fig:rounds-je}.

%----------------------------------------------------------------------
\subsection{Preference complexity and reporting accuracy}
\label{sec:misreporting}

We use two measures of reporting accuracy. The first, \emph{choice accuracy}, evaluates whether a participant ultimately receives her most-preferred program (by the \emph{true} ranking) among the seats available at her turn. This measure enables a fair comparison between direct mechanisms and SD-CHOICE. In ACCURACY, there is no allocation; to create comparable choice sets, we simulate Random Priority markets using participants’ reported preferences: in each market we draw a random priority order without replacement and allocate by serial dictatorship, repeating this 100 times and averaging at the market level.\footnote{For each market of 27 participants, we run 100 independent simulations. In each, we draw a unique priority order (1–27) without replacement and allocate via Random Priority under reported preferences. For a given participant and simulation, choice accuracy equals 1 if her top reported program among the available seats at her turn matches her top true program among those same available seats. We then average within market.} Formally, choice accuracy equals one if the participant’s top-ranked reported program among the available set coincides with her top-ranked true program among that set.

A potential concern is that choice accuracy may downweight errors made by low-priority participants. We therefore also report a cardinal measure of deviation from truthful reporting that ignores priority: the normalized Kendall distance between the submitted and true rankings over all 27 items, taking values in $[0,1]$ (0 = identical; 1 = complete reversal).\footnote{An alternative and more typical measure would be a dummy for submitting the exact truthful ranked list. In typical experiments with short lists (e.g., up to eight items; see \citet{Hakimov2021a}) this is informative. In our setting with 27 items, exact truth-telling is rare, so a dummy would discard too much information; moreover, by design it would always equal zero in SD-WEIGHT and SD-LEX under SEP and COMP, where the interfaces cannot express the exact truthful ranking.} Note that Kendall distance cannot be calculated in SD-CHOICE, as we observe only one choice per participant.

%------------------------------
% Table: Summary Table 1
%------------------------------
\begin{table}[h]
    \centering
    \begin{threeparttable}
    \caption{Summary Table: Reporting Accuracy}
    \label{tab:accuracy-complexity}
    \begin{tabular}{l C{1.25cm} C{1.25cm} C{1.25cm} C{1.25cm} C{1.25cm} C{1.25cm} C{1.25cm} C{1.25cm}}
        \toprule
        \multirow{2}{*}{Treatments} 
        & \multicolumn{4}{c}{Choice Accuracy (\%)} 
        & \multicolumn{4}{c}{Kendall Distance} \\
        \cmidrule(r){2-5} \cmidrule(l){6-9}
        & Overall & LEX & SEP & COMP & Overall & LEX & SEP & COMP \\
        \midrule
        ACCURACY  & 76.72 & 92.00 & 67.77 & 70.38 & 0.11 & 0.04 & 0.13 & 0.17 \\
        SD-DIRECT & 55.76 & 63.89 & 49.54 & 53.86 & 0.27 & 0.22 & 0.29 & 0.29 \\
        SD-WEIGHT & 49.85 & 57.56 & 46.76 & 45.22 & 0.28 & 0.23 & 0.26 & 0.33 \\
        SD-LEX    & 47.84 & 65.43 & 40.28 & 37.81 & 0.33 & 0.21 & 0.37 & 0.41 \\
        SD-CHOICE & 84.26 & 91.67 & 85.65 & 75.46 & $\times$ & $\times$ & $\times$ & $\times$ \\
        \bottomrule
    \end{tabular}
    \begin{tablenotes}[flushleft]\footnotesize\parindent=1em
    \item \textit{Notes:} 
    Choice Accuracy is the market-level average share of participants who, at their turn, choose (or whose submitted ranking selects) the truly best program among those still available; 27 participants per market; 72 markets per treatment.
    Kendall Distance is the average normalized Kendall (tau) distance between a participant’s reported and true rankings. Kendall distance is not defined for SD-CHOICE because no full ranking is elicited.
    \end{tablenotes}
    \end{threeparttable}
\end{table}

As Table~\ref{tab:accuracy-complexity} shows (columns 2–5 under “Choice Accuracy”), LEX preferences yield the highest accuracy across all treatments, with differences relative to SEP and COMP significant at $p<0.01$ (two-treatment-subsample regressions, untabulated). For example, in ACCURACY, choice accuracy is 92.00\% in LEX but only 67.77\% in SEP and 70.38\% in COMP. The same pattern appears in Kendall distance: in every treatment where Kendall distance is defined, LEX has the lowest distance, significantly below SEP and COMP ($p<0.01$, untabulated). Nonparametric tests at the market level (Kruskal–Wallis) also reject equality across preference domains ($p<0.01$, untabulated).

By contrast, the ranking between SEP and COMP is not systematic and depends on the reporting interface. Choice accuracy and Kendall distance occasionally disagree on which of SEP or COMP performs better. Overall, we do not reject equality between SEP and COMP.

Table~\ref{accuracyreg} confirms these patterns. In columns (2) and (4), the COMP dummy is sizeable and highly significant ($p<0.01$), indicating larger Kendall distances and a lower probability of choosing the correct program. LEX, the simplest domain, consistently yields the highest accuracy (and lowest Kendall distance). There is no significant difference between SEP and COMP coefficients.

Thus, \textsc{Hypothesis 1} is supported: LEX generates the most accurate reporting in every treatment. Between SEP and COMP---the comparison we flagged as exploratory---we find no robust difference: SEP already imposes sufficient computational burden to trigger substantial errors, and participants appear to absorb COMP's additional complexity through effort rather than accuracy.\footnote{LEX is the only domain that can often be resolved without extensive calculator use or cross-attribute computations. Under SEP and COMP, subjects must rely on on-screen calculations and documentation of scores for many programs; this likely depresses accuracy in both domains. While COMP is theoretically more complex, the calculator may attenuate the incremental difficulty relative to SEP, yielding similar accuracy.}

Response times corroborate this interpretation. In ACCURACY, participants spend on average 270 seconds per round in LEX but around 460 and 445 seconds in SEP and COMP, and the 8-minute limit binds precisely where complexity is high: 59\% of SEP rounds and 50\% of COMP rounds use at least 470 of the 480 available seconds, against 6\% in LEX. Time use in SEP and COMP is thus censored at the limit, consistent with errors plateauing between the two domains: participants appear to exhaust their effort capacity already in SEP. In SD-CHOICE, where the per-decision limit rarely binds, decision times are monotone in complexity---64, 101, and 115 seconds per pick in LEX, SEP, and COMP---confirming the intended complexity ranking behaviorally.

The nature of the errors points to heuristic simplification rather than pure calculation noise. Call a pair of programs \emph{dominance-ordered} if the two share the field of study and one is weakly better on both remaining attributes (higher university prestige, lower tuition). Ranking the dominated program higher is an unambiguous mistake whenever the induced utility agrees with dominance, as it always does under LEX and SEP. In ACCURACY, the share of submitted lists containing at least one such violation rises from 8\% in LEX to 52\% in SEP and 80\% in COMP. Under COMP, moreover, the interaction term reverses surface dominance for many pairs (about 35 per list), so that the truly better program \emph{looks} worse attribute by attribute; participants side with surface dominance---against their induced preferences---in 35\% of these pairs, compared with a 6\% error rate on pairs where dominance and utility align. Attribute interactions are thus systematically underweighted. Consistent with simplification, in SEP the submitted rankings are on average closer to the space of lexicographic orderings than the true rankings are (mean minimum Kendall distance of 0.12 versus 0.16): when trade-offs are hard, participants tilt toward attribute hierarchies.

\begin{result}[Preference complexity]
\label{res:comp-complexity}
Reports are more accurate in LEX than in SEP or COMP: choice accuracy is higher and Kendall distance lower, in every treatment. We do not find a robust difference between SEP and COMP.
\end{result}

\begin{table}[htbp]\centering
\def\sym#1{\ifmmode^{#1}\else\(^{#1}\)\fi}
\caption{Choice Accuracy and Kendall Distance Regressions}
\label{accuracyreg}
\setlength{\tabcolsep}{15pt} % 调整列间距
\begin{threeparttable}
\begin{tabular}{l*{5}{c}}
\toprule
 & \multicolumn{2}{c}{Choice Accuracy} & \multicolumn{2}{c}{Kendall Distance} \\
\cmidrule(lr){2-3} \cmidrule(lr){4-5} % 局部横线
 & (1) & (2) & \multicolumn{1}{c}{(3)} & \multicolumn{1}{c}{(4)} \\
\midrule
\textit{Treatment:}                &                     &                     &                     &                     \\
SD-DIRECT             &      -0.207\sym{***}&      -0.207\sym{***}&       0.151\sym{***}&       0.151\sym{***}\\
                    &     (0.023)         &     (0.023)         &     (0.013)         &     (0.013)         \\
SD-WEIGHT              &      -0.267\sym{***}&      -0.267\sym{***}&       0.163\sym{***}&       0.163\sym{***}\\
                    &     (0.013)         &     (0.013)         &     (0.009)         &     (0.009)         \\
SD-LEX                 &      -0.288\sym{***}&      -0.288\sym{***}&       0.217\sym{***}&       0.217\sym{***}\\
                    &     (0.012)         &     (0.012)         &     (0.008)         &     (0.008)         \\
SD-CHOICE                   &       0.075\sym{***}&       0.075\sym{***}&                     &                     \\
                    &     (0.014)         &     (0.014)         &                     &                     \\
\textit{Pref. Type:}                &                     &                     &                     &                     \\
SEP                  &                     &      -0.169\sym{***}&                     &       0.090\sym{***}\\
                    &                     &     (0.017)         &                     &     (0.011)         \\
COMP                &                     &      -0.176\sym{***}&                     &       0.127\sym{***}\\
                    &                     &     (0.017)         &                     &     (0.012)         \\
Constant            &                     &                     &       0.114\sym{***}&       0.042\sym{***}\\
                    &                     &                     &     (0.006)         &     (0.009)         \\
\midrule
Observations        &        9720         &        9720         &        7774         &        7774         \\
\bottomrule
\end{tabular}
\begin{tablenotes}[flushleft]\footnotesize\parindent=0em
\item Notes: Mixed-effects regressions. Choice Accuracy is a dummy that equals 1 if a participant selected the truly best program among those remaining after higher-priority participants have made their selections, and 0 otherwise. Kendall Distance represents the proportion of discordant pairs in a participant’s reported ranking list compared to her true preferences, ranging from 0 to 1. Choice accuracy effects are expressed as marginal effects from logistic mixed-effects regressions. The baseline preference type is LEX. The baseline treatment is ACCURACY. For the ACCURACY treatment, data from only the first simulation are used (results are unchanged when averaging over simulations). Kendall-distance regressions contain 7,774 rather than 7,776 observations because two first-round SD-DIRECT reports were invalid submissions (placeholder or duplicate entries only), leaving the Kendall distance undefined. Standard errors in parentheses are clustered by matching group, {*} $\textit{p} < 0.1$, {*}{*} $\textit{p} < 0.05$, {*}{*}{*} $\textit{p} < 0.01$.
\end{tablenotes}
\end{threeparttable}
\end{table}

\subsection{Additional reporting errors under direct serial dictatorship}
\label{sec:accvsrp}

We compare ACCURACY to SD-DIRECT. The two treatments hold fixed the full-ranking task and the induced preferences; they differ in the presence of allocation incentives and in the payoff rule: ACCURACY directly rewards agreement with the induced ranking, whereas SD-DIRECT pays through the assignment outcome. We therefore interpret the ACCURACY--SD-DIRECT gap as an incentive-and-framing gap---a composite of incentive-related forces such as misperceived incentives, reference dependence, rank utility, and disappointment aversion---rather than as a pure estimate of strategic misreporting. List-length burden and preference complexity, by contrast, are held fixed across the two treatments.

As shown in Table~\ref{tab:accuracy-complexity}, overall choice accuracy in SD-DIRECT is 55.76\% versus 76.72\% in ACCURACY, a 21 percentage point difference. The gap is significant overall and within each preference domain ($p<0.01$). Thus, while complexity of reporting contributes to errors in both treatments, SD-DIRECT exhibits substantial additional misreporting consistent with incentive-related misreporting under Serial Dictatorship.

\begin{figure}[!ht]
    \centering
    \includegraphics[width=0.95\linewidth]{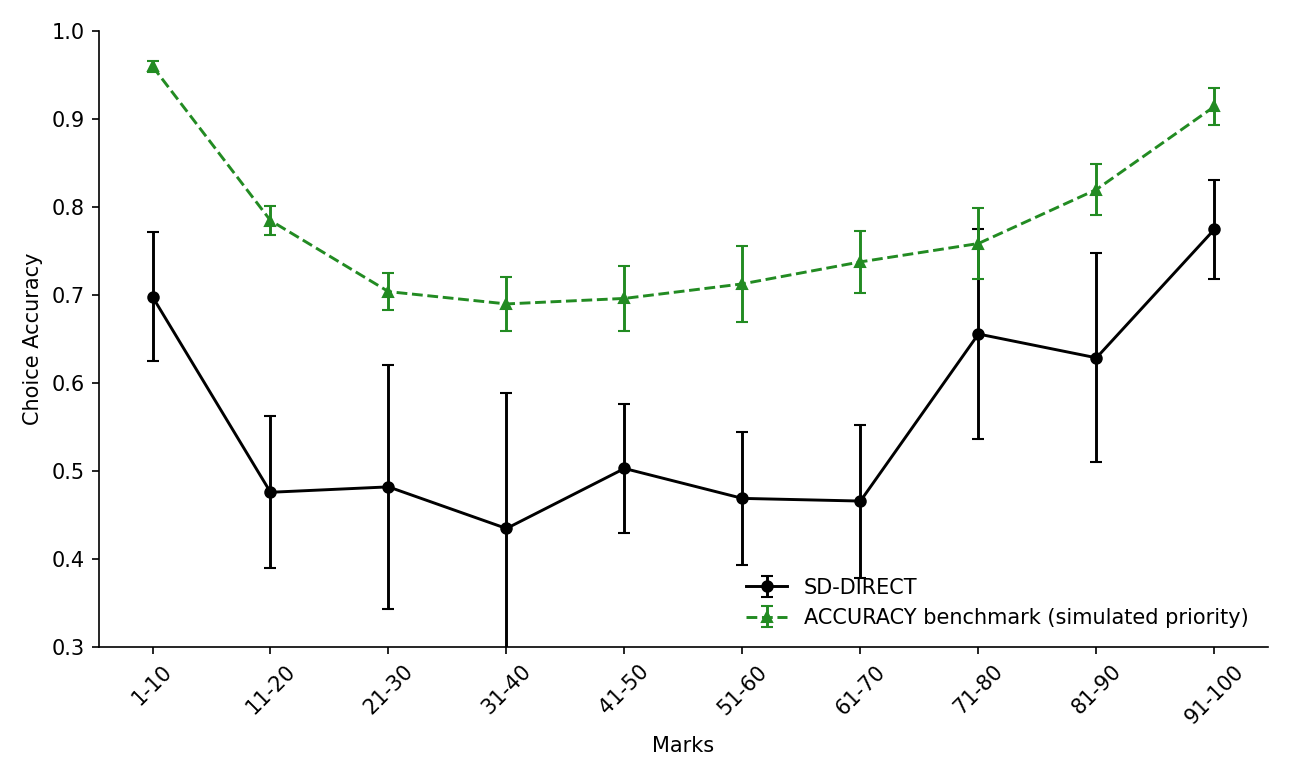}
    \caption{The Role of Marks}
    \label{fig:marks}
    \par
    \footnotesize{Notes: The groups (e.g., “1–10”, “91–100”) refer to the marks assigned to participants. The ACCURACY benchmark evaluates the incentive-free ACCURACY reports position by position, using the 100 simulated random-priority markets of Table~\ref{tab:accuracy-complexity}; each simulated priority position is assigned to the mark bin with the corresponding expected mark. CIs at 95\%, based on matching-group means.}
\end{figure}

Further evidence comes from the gradient of accuracy by marks in SD-DIRECT, presented in Figure~\ref{fig:marks}. Choice accuracy is U-shaped: around 77\% for the highest marks (91--100), around 70\% for the lowest (1--10), and 44--50\% for intermediate marks. Part of this U-shape, however, is mechanical rather than incentive-driven. The green series in Figure~\ref{fig:marks} evaluates the incentive-free ACCURACY reports position by position, using the same simulated random-priority markets as in Table~\ref{tab:accuracy-complexity}, and it is U-shaped as well. The intuition is simple: choice accuracy at priority position $p$ probes the ordering of a report around depth $p$ of the list. The top and bottom of a 27-item list are comparatively easy to order, because adjacent programs there are far apart in score, whereas mid-list programs are bunched together and hard to order exactly; and at the last positions only a few programs remain, leaving few ways to err. A report submitted with no knowledge of one's position therefore scores high at extreme positions and lower in the middle, absent any strategic motive. The incentive-and-framing component of Figure~\ref{fig:marks} is the \emph{excess} shortfall of SD-DIRECT below this benchmark: it averages about 14 percentage points in the top three mark bins---where truthful ranking of the top few programs is both simple and evidently sufficient---and about 25 percentage points everywhere else. Notably, participants with the lowest marks fall furthest below their mechanical potential (70\% against a 96\% benchmark), even though their task is the easiest: identifying the best among the few remaining programs.\footnote{This pattern aligns with \citet{hassidim2021limits}, who show that applicants with poor grades are more likely to submit dominated rank-order lists than those with better grades.} The benchmark also reinforces the caveat that ACCURACY is not a frictionless ideal: even absent strategic incentives, reporting frictions interact with the allocation mechanics in a systematic, position-dependent way---which is why we interpret ACCURACY throughout as a measure of reporting frictions rather than of perfect truth-telling.

These findings support \textsc{Hypothesis 2}. Mixed-effects regressions in Table~\ref{accuracyreg} corroborate the raw differences: the SD-DIRECT indicator is associated with significantly lower choice accuracy and higher Kendall distance relative to ACCURACY (all $p<0.01$).

\begin{result}[Additional reporting errors under direct serial dictatorship]
\label{res:accvssd}
Within each domain (LEX, SEP, COMP), reports are less accurate in SD-DIRECT than in ACCURACY: choice accuracy is significantly lower and Kendall distance substantially higher. The gap is consistent with incentive-related and framing-related reporting errors under serial dictatorship.
\end{result}

\subsection{Reporting interfaces and accuracy}
\label{sec:interfaces}

We next compare the three direct reporting interfaces: SD-DIRECT (a full ranking over 27 programs), SD-WEIGHT (separate rankings over the three attributes plus attribute weights), and SD-LEX (separate attribute rankings aggregated lexicographically). Behavioral Hypothesis~3 predicts that the interfaces' performance relative to SD-DIRECT deteriorates with their expressiveness loss---strongly for SD-LEX, mildly for SD-WEIGHT---while on LEX the sign of the comparison depends on whether the reduced input burden outweighs interface-use costs. 

In the LEX domain, all three interfaces are fully expressive (Table~\ref{tab:expressiveness}); SD-WEIGHT expresses lexicographic orderings exactly when weights are sufficiently separated. In the data (Table~\ref{tab:accuracy-complexity}, LEX columns), SD-LEX attains the highest choice accuracy and the lowest Kendall distance among the three, but its advantage over SD-DIRECT is not statistically significant ($p>0.10$); by contrast, SD-WEIGHT performs significantly worse than SD-LEX ($p<0.01$) and worse than SD-DIRECT ($p<0.05$). Hence we do not find a clear accuracy gain from simplified reporting even when the interface matches the preference domain. 

When preferences require trade-offs (SEP) or include attribute interactions (COMP), SD-DIRECT yields the highest choice accuracy; the advantage is significant relative to SD-LEX in SEP and significant relative to both SD-LEX and SD-WEIGHT in COMP ($p<0.01$). As complexity rises, SD-WEIGHT delivers higher choice accuracy than SD-LEX ($p<0.01$), consistent with weights partially offsetting lexicographic misspecification, and this ordering is mirrored by Kendall distance. Notably, SD-WEIGHT outperforms SD-DIRECT in SEP on Kendall distance ($p<0.05$, untabulated), though this advantage does not translate into higher choice accuracy. 

Aggregating across domains, neither SD-LEX nor SD-WEIGHT improves accuracy relative to SD-DIRECT; SD-DIRECT achieves the highest overall choice accuracy and the lowest Kendall distance. Among direct mechanisms in our environment, the full-ranking interface is thus more robust than the two attribute-based restrictions. This cautions against adopting lexicographic or weighted-attribute reporting formats solely because they reduce the length of the submitted message.

These patterns are consistent with \textsc{Hypothesis 3}. SD-LEX's relative performance collapses from $+1.5$ percentage points in LEX to $-9.3$ and $-16.1$ in SEP and COMP, tracking its expressiveness floors, and it falls behind SD-WEIGHT in both non-lexicographic domains; SD-WEIGHT's relative performance varies far less ($-6.3$, $-2.8$, and $-8.6$). The LEX comparison resolves the ex-ante ambiguity: the reduced input burden does not outweigh interface-use costs.

\begin{result}[Reporting interfaces]
\label{res:interfaces}
Relative to SD-DIRECT, neither SD-WEIGHT nor SD-LEX significantly increases accuracy—even in LEX. Aggregating across preference domains, SD-DIRECT yields the highest choice accuracy and the lowest Kendall distance; the choice-accuracy differences relative to SD-LEX and SD-WEIGHT are significant. Relative performance deteriorates with an interface's expressiveness loss, in line with Hypothesis~3.
\end{result}

\paragraph{Decomposing the interface penalty.}
How does each restricted interface's Kendall distance compare with SD-DIRECT's, and how much of the difference is mechanical? For each SD-WEIGHT and SD-LEX participant-round we compute the \emph{representational floor} $F_j$---the minimum normalized Kendall distance between her realized true ranking and any ranking the interface can express---and split the difference from SD-DIRECT into two terms:
\begin{equation*}
\underbrace{D_j - D_{\text{DIRECT}}}_{\text{net difference}}
\;=\; \underbrace{F_j}_{\text{representational floor}}
\;+\; \underbrace{\big[(D_j - F_j) - D_{\text{DIRECT}}\big]}_{\text{difference in distance above the floor}}.
\end{equation*}
The floor is the unavoidable distortion induced by the message space. The distance above the floor measures how far reports fall short of the best expressible ranking; because participants were paid through allocations rather than for minimizing Kendall distance, it is not a pure measure of ability to use the interface---it may include strategic adaptation to the restricted message space, imperfect optimization, misunderstanding, translation error, and noise. Table~\ref{tab:decomposition} reports the decomposition by interface and preference domain, computing each floor from the participant-round's realized preference profile.

\begin{table}[h]
\centering
\caption{Kendall Distance Relative to SD-DIRECT: Representational Floors and Distance Above the Floor}
\label{tab:decomposition}
\begin{tabular}{llrrrrr}
\toprule
Interface & Domain & Observed & Floor & Above floor & SD-DIRECT & Net difference \\
 & & $D_j$ & $F_j$ & $D_j-F_j$ & $D_{\text{DIRECT}}$ & $D_j-D_{\text{DIRECT}}$ \\
\midrule
SD-WEIGHT  & LEX  & 0.234 & 0.000 & 0.234 & 0.216 & $+0.018$ \\
SD-WEIGHT  & SEP  & 0.265 & 0.024 & 0.241 & 0.288 & $-0.023$ \\
SD-WEIGHT  & COMP & 0.331 & 0.088 & 0.242 & 0.291 & $+0.039$ \\
SD-LEX     & LEX  & 0.211 & 0.000 & 0.211 & 0.216 & $-0.005$ \\
SD-LEX     & SEP  & 0.371 & 0.231 & 0.140 & 0.288 & $+0.083$ \\
SD-LEX     & COMP & 0.410 & 0.295 & 0.115 & 0.291 & $+0.119$ \\
\bottomrule
\end{tabular}

\vspace{0.5em}
\begin{minipage}{0.95\textwidth}
\footnotesize \textit{Notes:} ``Observed'' is the average normalized Kendall distance between submitted and true rankings. ``Floor'' is the average of participant-round-specific representational floors, computed from each realized preference profile against the enumerated message space of Section~\ref{sec:expressiveness}. ``Above floor'' is Observed minus Floor. ``Net difference'' is Observed minus the SD-DIRECT average in the same domain and equals Floor plus the difference in distance above the floor.
\end{minipage}
\end{table}

The decomposition reveals a sharper pattern than a simple usability failure. In SEP and COMP, \emph{both} restricted interfaces exhibit less distance above the floor than SD-DIRECT's total distance ($0.24$ vs.\ $0.29$ for SD-WEIGHT; $0.12$--$0.14$ vs.\ $0.29$ for SD-LEX): conditional on what the interface can express, restricted reports track the best expressible ranking comparatively well. Representational restrictions offset these gains---substantially for SD-LEX, whose floors of $0.23$--$0.30$ produce a net increase in Kendall distance of $0.08$--$0.12$, and more modestly for SD-WEIGHT, whose small floors leave its net difference close to zero ($-0.02$ in SEP, $+0.04$ in COMP). SD-WEIGHT's small net improvement in Kendall distance in SEP does not, however, translate into better assignment accuracy (Table~\ref{tab:accuracy-complexity}). On LEX, where both floors are zero, the two interfaces perform about as well as SD-DIRECT on Kendall distance.\footnote{Participants do adapt their weights to the domain in the right direction: median submitted weights (university/field/tuition) are $80/20/10$ in LEX rounds but $35/38.5/35$ in SEP rounds. Precision is imperfect: even in LEX, only 61\% of submissions satisfy the ordering $w_U > w_F > w_T$---itself only a necessary condition, weaker than the weight separation required to implement lexicographic preferences exactly.}

This decomposition sharpens Result~\ref{res:interfaces}: restricted interfaces reduce ranking error conditional on what they can express, but these gains are offset by representational restrictions---substantially for SD-LEX and more modestly for SD-WEIGHT---and neither interface improves assignment accuracy.

\subsection{Sources of benefits of SD-CHOICE}
\label{sec:sd-choice-decomp}

Tables~\ref{tab:accuracy-complexity} and \ref{accuracyreg} show that SD-CHOICE delivers substantially higher choice accuracy than SD-DIRECT (84.26\% vs.\ 55.76\%), a 28.5 percentage-point gain ($p<0.01$). This replicates experimental findings of \citet{Li2017} and \citet{Bo2024} and is consistent with incentive strength: SD-CHOICE is obviously strategy-proof and ``1-step simple'' \citep{Pycia2023}, whereas SD-DIRECT is not. However, SD-CHOICE also simplifies the reporting interface: participants make a single pick from the current menu rather than constructing a full ranking over 27 items. The improvement therefore arises from two sources---a menu-size channel and everything else that distinguishes sequential choice, including obvious strategy-proofness and the one-at-a-time decision frame. In this subsection we quantify the menu-size component and treat the rest as a residual.

We decompose the headline effect directly, using one observation per participant and round in \emph{both} treatments and the same outcome as in Table~\ref{tab:accuracy-complexity}: the indicator $Y$ that the participant obtains her truly best program among those available at her turn. The two treatments differ in the menu relevant to this task. In SD-CHOICE, a participant picks from the $m$ programs remaining at her realized priority position. In SD-DIRECT, she constructs her report facing the full set of 27 programs---the mechanism removes nothing from her consideration set while she reports---so we set $m=27$. The randomized priority draw generates variation in $m$ within SD-CHOICE that traces out the position/menu gradient, which is what makes the decomposition feasible.

Let $s\in\{0,1\}$ be an indicator for SD-CHOICE. The ``short'' regression, estimated on the pooled SD-CHOICE and SD-DIRECT samples,
\begin{equation}
Y \;=\; \alpha_S + \delta_S\, s + \mathbf{x}'\boldsymbol{\gamma}_S + \varepsilon, 
\label{eq:short}
\end{equation}
where $\mathbf{x}$ contains preference-type fixed effects, recovers the total choice-accuracy gap, $\hat\delta_S$. The ``long'' regression adds a quadratic in menu size,
\begin{equation}
Y \;=\; \alpha_L + \delta_L\, s + \beta_1\, m + \beta_2\, m^2 + \mathbf{x}'\boldsymbol{\gamma}_L + u,
\label{eq:long}
\end{equation}
and we interpret $\hat\delta_L$ as the \emph{residual SD-CHOICE component}: the advantage that survives after netting out a flexible quadratic in menu size. The remainder is the \emph{menu-size component} and admits the closed-form Gelbach \citep{gelbach2016} expression
\begin{equation}
\underbrace{\hat\delta_S}_{\text{Total gap}}
\;-\;
\underbrace{\hat\delta_L}_{\text{Residual component}}
\;=\;
\underbrace{\hat\beta_1\,\hat\gamma_m + \hat\beta_2\,\hat\gamma_{m^2}}_{\text{Menu-size component}},
\label{eq:gelbach}
\end{equation}
where $\hat\gamma_m$ and $\hat\gamma_{m^2}$ are the coefficients on $s$ in auxiliary regressions of $m$ and $m^2$ on $s$ and $\mathbf{x}$. The menu-size component answers the counterfactual: \emph{how much of the SD-CHOICE advantage is explained by the fact that SD-CHOICE participants choose from menus smaller than the full set screened in SD-DIRECT?} The residual SD-CHOICE component captures everything else. We interpret it as including obvious strategy-proofness and the simpler one-at-a-time decision frame, but it should not be read as a pure estimate of either channel separately.\footnote{Because treatment is assigned at the session (matching-group) level, standard errors for all three components are obtained by a cluster bootstrap at the matching-group level with 500 replications \citep{cameron2015}, re-estimating equations~\eqref{eq:short}--\eqref{eq:gelbach} on each replicate. A participant-cluster bootstrap yields very similar standard errors.}

Because priority position is randomly assigned, the decomposition uses exogenous variation in position. Position, however, jointly determines the size and the composition of the remaining menu. We therefore interpret the estimated component as a reduced-form priority-position/menu component, rather than as the causal effect of menu size holding menu composition fixed. In the fitted model, the component corresponds to evaluating all SD-CHOICE observations at the full menu of 27 programs: predicted average choice accuracy in SD-CHOICE falls from $0.843$ to $0.716$, a drop of $0.127$. A nonparametric version that replaces the quadratic prediction with observed SD-CHOICE accuracy on nearly full menus ($m\in\{25,26,27\}$: $0.747$) yields $0.096$. Relative to the headline gap of $0.285$, the component is thus about one-third (nonparametric) to $45\%$ (quadratic).

\begin{figure}[t]
\centering
\includegraphics[width=0.85\textwidth]{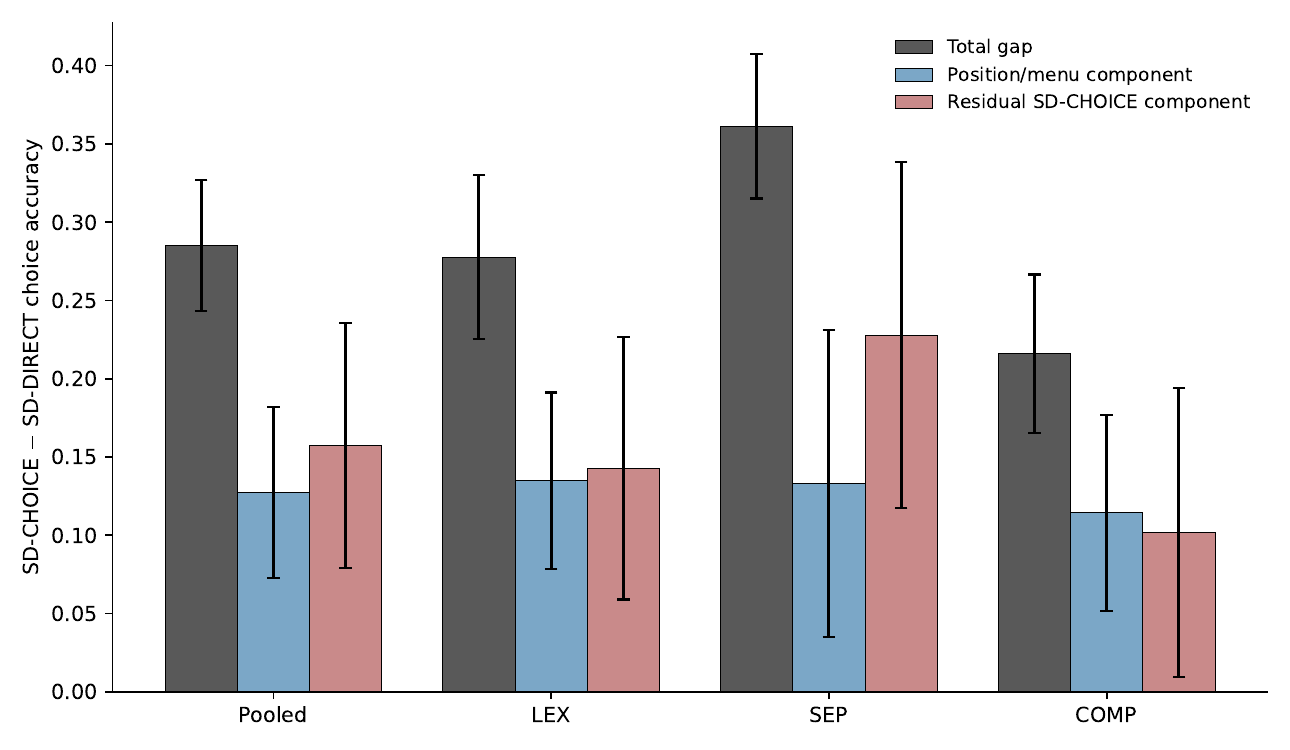}
\caption{Gelbach decomposition of the SD-CHOICE--SD-DIRECT gap}
\label{fig:decomp}
    \par
    \parbox[t]{\linewidth}{\raggedright
        \footnotesize{Notes: Gelbach decomposition of the SD-CHOICE--SD-DIRECT choice-accuracy gap into a position/menu component and a residual SD-CHOICE component, pooled and by preference domain. Whiskers show 95\% confidence intervals from a cluster bootstrap at the matching-group level, the level of treatment assignment.}
    }
\end{figure}

Table~\ref{tab:decomp} reports the decomposition pooled and by preference domain, and Figure~\ref{fig:decomp} displays the same information visually. Pooling across domains, the total SD-CHOICE--SD-DIRECT choice-accuracy gap is $0.285$---the headline effect of Table~\ref{tab:accuracy-complexity}. Of this, the position/menu component accounts for $0.127$ (about $45\%$), while the residual component accounts for $0.157$ (about $55\%$); both are significantly positive ($p<0.01$). The position/menu component is stable across domains ($0.11$--$0.14$ in LEX, SEP, and COMP), indicating that the benefit of acting on smaller, sequentially revealed menus is comparable across preference structures. The residual component varies more: $0.143$ (LEX), $0.228$ (SEP), and $0.102$ (COMP). It is largest in SEP, where preferences are nontrivial to form but still tractable, leaving substantial room for incentive-related and framing-related errors in SD-DIRECT to depress accuracy. In COMP, the cognitive burden of forming preferences is highest---accuracy is depressed in both SD-DIRECT and SD-CHOICE---compressing the margin on which the residual component can operate. Consistent with Result~\ref{res:comp-complexity}, this pattern echoes the absence of a robust raw-accuracy difference between SEP and COMP: COMP's additional cognitive burden eats into the residual gain that SD-CHOICE would otherwise deliver.

\begin{table}[t]
\centering
\caption{Gelbach decomposition of the SD-CHOICE--SD-DIRECT gap}
\label{tab:decomp}
\begin{tabular}{lccc}
\toprule
 & Total gap & Position/menu & Residual component \\
 & $\hat\delta_S$ & $\hat\beta_1\hat\gamma_m+\hat\beta_2\hat\gamma_{m^2}$ & $\hat\delta_L$ \\
\midrule
Pooled & 0.285 & 0.127 & 0.157 \\
LEX    & 0.278 & 0.135 & 0.143 \\
SEP    & 0.361 & 0.133 & 0.228 \\
COMP   & 0.216 & 0.114 & 0.102 \\
\bottomrule
\end{tabular}
\par\smallskip
\begin{minipage}{0.92\textwidth}
\footnotesize
\textit{Notes:} The dependent variable is choice accuracy as in Table~\ref{tab:accuracy-complexity}: an indicator that the participant obtains her truly best program among those available at her turn, with one observation per participant and round in both treatments. The total gap is the coefficient $\hat\delta_S$ on the SD-CHOICE indicator in the short regression~\eqref{eq:short}; the residual SD-CHOICE component is the analogous coefficient $\hat\delta_L$ in the long regression~\eqref{eq:long}, which adds a quadratic in menu size ($m$ = realized available-set size in SD-CHOICE; $m=27$, the full reporting menu, in SD-DIRECT); the position/menu component is the closed-form difference $\hat\delta_S-\hat\delta_L$, equal to $\hat\beta_1\hat\gamma_m+\hat\beta_2\hat\gamma_{m^2}$ where $\hat\gamma_m,\hat\gamma_{m^2}$ are the SD-CHOICE coefficients in auxiliary regressions of $m$ and $m^2$ on the SD-CHOICE indicator and preference-type fixed effects \citep{gelbach2016}. All regressions include preference-type fixed effects in the pooled row. Confidence intervals are from a cluster bootstrap at the matching-group level---the level of treatment assignment---with participant-level clustering as robustness (see Figure~\ref{fig:decomp}).
\end{minipage}
\end{table}

\begin{figure}[!ht]
    \centering
    \includegraphics[width=1\linewidth]{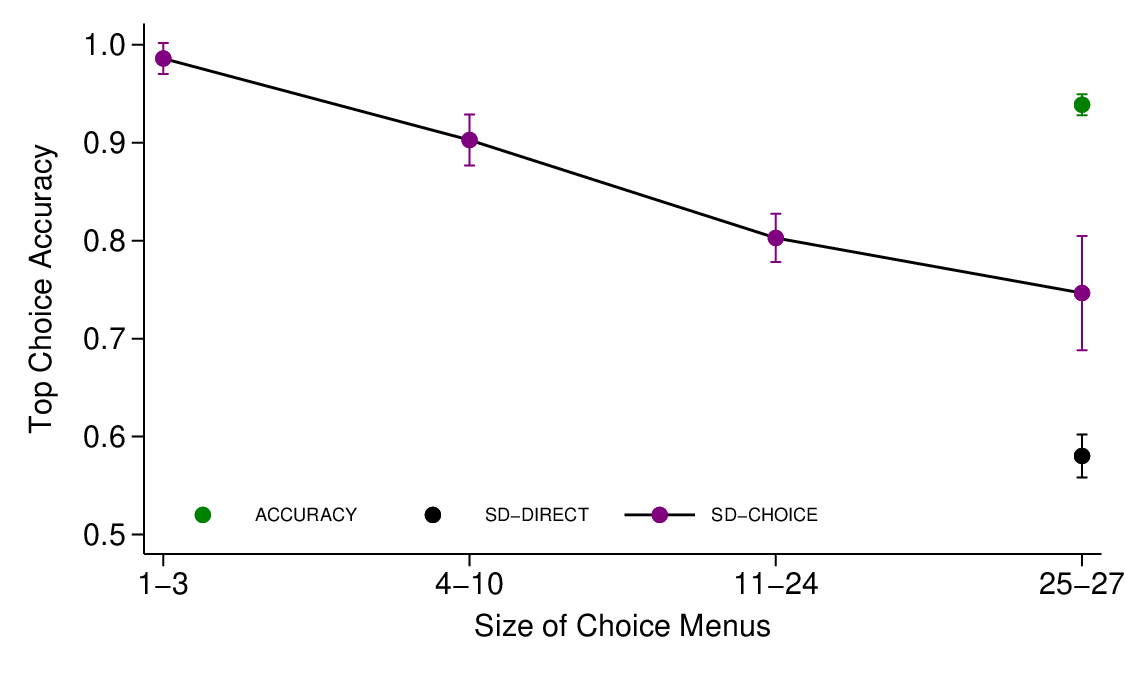}
    \caption{Choice Accuracy and Menu Size}
    \label{fig:size_ChoiceMenus}
    \par
    \parbox[t]{\linewidth}{\raggedright
        \footnotesize{Notes: The figure reports top-choice accuracy by the size of the choice menu. “Top-choice accuracy” indicates whether a subject correctly selected or ranked her truly most-preferred item first when making a choice or submitting a full ranking. For SD-CHOICE, the menu is the set of programs remaining at the participant's turn; ACCURACY and SD-DIRECT rankings are evaluated on the full menu of 27 programs.}
    }
\end{figure}

The decomposition attributes the residual $\hat\delta_L$ to a combination of obvious strategy-proofness and a simpler decision frame relative to ranking 27 items ex ante. Consistent with this, SD-CHOICE also exceeds the ACCURACY benchmark overall ($84.26\%$ vs.\ $76.72\%$, $p<0.01$), a comparison that holds preferences fixed and removes \emph{all} strategic incentives.\footnote{The SD-CHOICE--ACCURACY gap is largest and statistically significant in SEP ($p<0.01$), while differences in LEX and COMP are not significant; the latter comparisons depend on the simulated priority orders in ACCURACY.} Because ACCURACY still requires a full 27-item ranking, the SD-CHOICE--ACCURACY gap is consistent with an additional interface channel beyond incentive strength alone. Figure~\ref{fig:size_ChoiceMenus} corroborates this: in SD-CHOICE, choice accuracy rises sharply as menus shrink, while ACCURACY top-choice accuracy on the largest menu ($m\in\{25,26,27\}$) sits significantly above SD-CHOICE accuracy on the same menu---consistent with an important role for menu reduction in the SD-CHOICE advantage. The full-menu comparison is informative but not a pure incentive contrast: ACCURACY participants rank their truly best program first $94\%$ of the time, against $58\%$ at the top slot of SD-DIRECT, while SD-CHOICE picks on nearly full menus are accurate $75\%$ of the time. Because ACCURACY directly rewards ranking accuracy whereas SD-DIRECT pays through the allocation, the large ACCURACY--SD-DIRECT difference is consistent with incentive-related distortions under SD-DIRECT, but may also partly reflect the stronger accuracy framing in ACCURACY. Taken together, these findings support \textsc{Hypothesis 4}: SD-CHOICE improves accuracy relative to SD-DIRECT. The decomposition further suggests---as a post-hoc mechanism analysis---that part of the improvement reflects the simpler, one-at-a-time reporting interface rather than incentives alone.

\begin{result}[Sequential choice and menu size]
SD-CHOICE substantially increases accuracy relative to SD-DIRECT. In a reduced-form decomposition of the headline choice-accuracy effect, the priority-position/menu component accounts for $0.096$ to $0.127$ of the $0.285$ SD-CHOICE--SD-DIRECT gap, or roughly one-third to $45\%$. The remaining gap is a residual SD-CHOICE component that includes obvious strategy-proofness and the simpler one-at-a-time decision frame. SD-CHOICE also exceeds ACCURACY overall, with the clearest difference in SEP, suggesting that avoiding full-list construction can be behaviorally important.
\end{result}

\subsection{Efficiency loss and justified envy}
\label{sec:efficiencyenvy}

\begin{table}[h]
    \centering
    \begin{threeparttable}
    \caption{Summary Table: Efficiency Loss and Justified Envy}
    \label{tab:allocation-efficiency}
    \begin{tabular}{l C{1.25cm} C{1.25cm} C{1.25cm} C{1.25cm} C{1.25cm} C{1.25cm} C{1.25cm} C{1.25cm}}
        \toprule
        \multirow{2}{*}{Treatments} 
        & \multicolumn{4}{c}{Efficiency Loss (\%)} 
        & \multicolumn{4}{c}{Justified Envy (\%)} \\
        \cmidrule(r){2-5} \cmidrule(l){6-9}
        & Overall & LEX & SEP & COMP & Overall & LEX & SEP & COMP \\
        \midrule
ACCURACY  & 3.33 & 0.44 & 3.78 & 5.76  & 23.26 & 7.88 & 32.38 & 29.54 \\
SD-DIRECT  & 7.77 & 3.73 & 8.35 & 11.23 & 43.88 & 35.34 & 50.15 & 46.14 \\
SD-WEIGHT & 7.73 & 3.98 & 7.79 & 11.44 & 50.15 & 42.44 & 53.24 & 54.78 \\
SD-LEX        & 8.00 & 3.05 & 7.27 & 13.67 & 52.16 & 34.57 & 59.72 & 62.19 \\
SD-CHOICE & 1.75 & 0.30 & 1.07 & 3.89 & 15.64 & 8.18 & 14.20 & 24.54 \\
        \bottomrule
    \end{tabular}
    \begin{tablenotes}[flushleft]\footnotesize\parindent=1em
    \item \textit{Notes:} 
    {Efficiency Loss} is the average percentage of welfare loss relative to truthful serial dictatorship at the market level, calculated by
    \[
        E = \frac{M - R}{M}\times100,
    \]
    where $M$ is the welfare that serial dictatorship would deliver under the realized priority order if all participants reported truthfully, and $R$ is the realized welfare. The loss thus measures the cost of misreporting, holding the allocation rule and priorities fixed, rather than distance from a utilitarian optimum.
    {Justified Envy} reports the average proportion of participants exhibiting justified envy. A participant exhibits justified envy if another participant with a lower priority is assigned a seat that the envious participant prefers over her own assigned seat.
    \end{tablenotes}
    \end{threeparttable}
\end{table}

Reporting failures affect not only individual payoffs but also the aggregate outcome. To quantify these, we measure: (i) \emph{efficiency loss}, the shortfall of realized welfare from what serial dictatorship would deliver under the realized priority order if all participants reported truthfully (computed at the market level), and (ii) \emph{justified envy}, the fraction of participants who end up preferring someone else’s allocated seat despite having higher priority. Both measures thus capture the cost of misreporting under the allocation rule in place, not distance from a utilitarian optimum.

\begin{table}[htbp]\centering
\def\sym#1{\ifmmode^{#1}\else\(^{#1}\)\fi}
\caption{Efficiency Loss and Justified Envy Regressions}
\label{efficiencyreg}
\setlength{\tabcolsep}{15pt} 
\begin{threeparttable}
\begin{tabular}{l*{5}{c}}
\toprule
 & \multicolumn{2}{c}{Efficiency Loss} & \multicolumn{2}{c}{Justified Envy} \\
\cmidrule(lr){2-3} \cmidrule(lr){4-5} 
 & (1) & (2) & \multicolumn{1}{c}{(3)} & \multicolumn{1}{c}{(4)} \\
\midrule
\textit{Treatment:}  &                     &                     &                     &                     \\
SD-DIRECT    &       0.044\sym{***}&       0.043\sym{***}&       0.203\sym{***}&       0.203\sym{***}\\
                    &     (0.006)         &     (0.007)         &     (0.023)         &     (0.023)         \\
SD-WEIGHT   &       0.044\sym{***}&       0.044\sym{***}&       0.267\sym{***}&       0.266\sym{***}\\
                    &     (0.005)         &     (0.005)         &     (0.013)         &     (0.013)         \\
SD-LEX     &       0.047\sym{***}&       0.047\sym{***}&       0.288\sym{***}&       0.288\sym{***}\\
                    &     (0.004)         &     (0.004)         &     (0.012)         &     (0.012)         \\
SD-CHOICE    &      -0.016\sym{***}&      -0.016\sym{***}&      -0.076\sym{***}&      -0.076\sym{***}\\
                    &     (0.004)         &     (0.003)         &     (0.014)         &     (0.014)         \\
\textit{Pref. Type:} &                     &                     &                     &                     \\
SEP       &                     &       0.035\sym{***}&                     &       0.170\sym{***}\\
                    &                     &     (0.003)         &                     &     (0.017)         \\
COMP       &                     &       0.070\sym{***}&                     &       0.178\sym{***}\\
                    &                     &     (0.006)         &                     &     (0.017)         \\
Constant            &       0.033\sym{***}&      -0.002         &                     &                     \\
                    &     (0.003)         &     (0.003)         &                     &                     \\
\midrule
Observations        &         354         &         354         &        9720         &        9720         \\
\bottomrule
\end{tabular}
\begin{tablenotes}[flushleft]\footnotesize\parindent=0em
\item Notes: Efficiency Loss represents the fraction of welfare loss relative to truthful serial dictatorship at the market level; columns (1)--(2) are market-level linear regressions, with ACCURACY outcomes averaged over 100 simulated random-priority allocations and six markets containing invalid submissions excluded. Justified Envy is a dummy that equals 1 if a participant exhibited justified envy, and 0 otherwise; columns (3)--(4) report average marginal effects from logistic mixed-effects regressions, using data from only the first ACCURACY simulation (results are unchanged when averaging over simulations). The baseline preference type is LEX. The baseline treatment is ACCURACY. Standard errors in parentheses are clustered by matching group, {*} $\textit{p} < 0.1$, {*}{*} $\textit{p} < 0.05$, {*}{*}{*} $\textit{p} < 0.01$.
\end{tablenotes}
\end{threeparttable}
\end{table}

We now show that the treatment differences documented above for choice accuracy and Kendall distance translate into economically meaningful differences in welfare and fairness. Table~\ref{tab:allocation-efficiency} reports efficiency loss (shortfall from the welfare achievable under truthful reporting) and the share of participants with justified envy (a higher-priority participant preferring a lower-priority participant’s assignment). Across treatments and within each preference domain, lower accuracy coincides with more justified envy; for efficiency loss, the association is driven by the large accuracy gaps between mechanism classes---ACCURACY and SD-CHOICE against the three direct interfaces---whose losses are similar to one another. Moving from LEX to SEP to COMP, efficiency loss rises in every treatment (e.g., under SD-DIRECT: 3.73\% $\rightarrow$ 8.35\% $\rightarrow$ 11.23\%; under ACCURACY: 0.44\% $\rightarrow$ 3.78\% $\rightarrow$ 5.76\%), indicating that misreporting in more complex domains is costlier.

Aggregating across domains, SD-CHOICE attains the lowest efficiency loss (1.75\% overall; 3.89\% in COMP) and the lowest justified envy (15.64\% overall), while SD-LEX performs worst on both metrics (overall justified envy above 50\%). The regressions in Table~\ref{efficiencyreg} confirm these patterns: relative to ACCURACY, SD-DIRECT, SD-LEX, and SD-WEIGHT significantly increase efficiency loss and justified envy (all $p<0.01$), whereas SD-CHOICE significantly reduces both (efficiency loss coefficient $-0.016$, $p<0.01$; justified envy coefficient $-0.076$, $p<0.01$). Exact randomization inference (Appendix Table~\ref{tab:ri}) supports the ACCURACY and SD-CHOICE comparisons at the smallest attainable $p$-value ($0.002$); efficiency-loss differences between SD-DIRECT and the two attribute-based interfaces are not statistically distinguishable. Preference complexity independently worsens outcomes: SEP and COMP enter positively and significantly in both specifications. Taken together, these results show that reporting errors occur in consequential parts of the preference lists and translate into lower realized welfare relative to truthful play and into more violations of priority fairness; sequential choice mitigates both by improving reporting.
 
\begin{result}[Efficiency and envy]
\label{res:efficiency-envy}
Treatment differences in reporting accuracy translate into meaningful priority violations and, across mechanism classes, into welfare losses. SD-CHOICE achieves significantly lower efficiency loss and justified envy than all other treatments, consistent with its accuracy advantage; efficiency losses among the three direct interfaces are similar.
\end{result}

\section{Conclusion}
\label{sec:conclusion}

This paper studies how preference complexity and the reporting interface shape reporting accuracy and, in turn, welfare and justified envy in multi-attribute matching. Three facts emerge. First, complexity matters: LEX yields the highest accuracy; SEP and COMP substantially reduce accuracy with no robust difference between them. Second, incentives and framing matter: holding the reporting task fixed, ACCURACY outperforms SD-DIRECT, consistent with incentive-related and framing-related reporting errors under Serial Dictatorship. Third, the interface matters: attribute-based simplifications (SD-LEX, SD-WEIGHT) do not improve accuracy relative to a full ranking—even in LEX—while a sequential implementation (SD-CHOICE) performs better on accuracy and, consequently, on efficiency and justified envy. A higher choice accuracy than in ACCURACY and a reduced-form decomposition that attributes roughly one-third to $45\%$ of the SD-CHOICE--SD-DIRECT choice-accuracy gap to a priority-position/menu component indicate that part of SD-CHOICE's advantage arises from reducing the cognitive burden of full-list construction, not only from stronger incentive properties.

The results have two implications for assignment design. First, reducing the length of a report need not improve behavior. The decomposition in Section~\ref{sec:interfaces} shows that restricted interfaces reduce ranking error conditional on what they can express, but these gains are offset by representational restrictions---substantially for SD-LEX and more modestly for SD-WEIGHT---and neither interface improves assignment accuracy. Because designers rarely know the distribution of preference domains ex ante, the full-ranking interface performs more robustly in our experiment. Second, sequential and staged procedures can improve performance by reducing the active menu as well as by simplifying incentives. These findings do not imply that full one-by-one implementation is always practical, but they suggest that menu reduction is a relevant design margin for large assignment platforms.

Centralized platforms may face heavy communication, timing, and coordination costs if every participant must act in serial order. In practice, dynamic implementations can approximate small menus without literal one-by-one moves. Multi-offer or staged procedures progressively resolve uncertainty about high-priority candidates’ choices, shrinking remaining menus for others. This de facto “menu pruning” is central to recent deployments and proposals in higher education: dynamic or hybrid designs that batch early offers and elicit updated preferences have been used in France \citep[e.g.,][]{Hakimov2023a}, Inner Mongolia \citep{Gong2024}, and Tunisia \citep{Luflade2017}, and related mechanisms show preference discovery benefits in university admissions \citep{Grenet2022}. Our results provide evidence on one mechanism through which such policies may operate: smaller, more informative menus reduce errors and improve both efficiency and priority fairness.

Finally, the cost of direct reporting remains substantial and is likely higher still in real-world applications with far larger choice sets. The attribute-based formats we study did not yield behavioral benefits, but this does not imply that better simplified interfaces cannot be designed. Future work should study reporting interfaces that preserve expressiveness while reducing the cost of constructing full rankings, including assisted or adaptive elicitation procedures (see, e.g., \citealp{soumalias2025llm}).

\newpage

\section*{Declarations}
\noindent\textbf{Ethics approval.} The experiment was approved by the LABEX Ethics Board of HEC Lausanne on 08.03.2024, approval no. SAFER.

\noindent\textbf{Preregistration.} The experiment was preregistered at the AEA RCT Registry (AEARCTR-0013135).

\noindent\textbf{Data and code availability.} The data and all analysis code will be included in the replication package.

\noindent\textbf{Conflicts of interest.} The authors declare no conflicts of interest.

\noindent\textbf{Funding.} Financial support is acknowledged on the title page (National Natural Science Foundation of China, Grant \#W2433170; Swiss National Science Foundation, Project \#10003543).

\bibliographystyle{ecca}
\bibliography{ref}

\newpage
\appendix

\section{Appendix: Supplementary Analysis}

\begin{table}[htbp]
\centering
% Define the symbol command for significance stars, as in your original file
\def\sym#1{\ifmmode^{#1}\else\(^{#1}\)\fi}

\caption{Pairwise Comparison of Treatment Effects on Choice Accuracy (Preference Type: Overall)}
\label{tab:pairwise}
\begin{tabular}{@{}lccccc@{}}
\toprule
& \multicolumn{5}{c}{\textbf{Comparison Treatment}} \\
\cmidrule(l){2-6}
\textbf{Base Treatment} & SD-WEIGHT & SD-LEX & SD-DIRECT & SD-CHOICE & ACCURACY \\
\midrule
\textbf{SD-WEIGHT} & --- & -0.080\sym{**} & 0.238\sym{***} & 1.684\sym{***} & 1.203\sym{***} \\
& & & & & \\
\textbf{SD-LEX} & 0.080\sym{**} & --- & 0.318\sym{***} & 1.764\sym{***} & 1.284\sym{***} \\
& & & & & \\
\textbf{SD-DIRECT} & -0.238\sym{***} & -0.318\sym{***} & --- & 1.446\sym{***} & 0.966\sym{***} \\
& & & & & \\
\textbf{SD-CHOICE} & -1.684\sym{***} & -1.764\sym{***} & -1.446\sym{***} & --- & -0.481\sym{***} \\
& & & & & \\
\textbf{ACCURACY} & -1.203\sym{***} & -1.284\sym{***} & -0.966\sym{***} & 0.481\sym{***} & --- \\
\bottomrule
\multicolumn{6}{p{0.9\textwidth}}{\footnotesize \textit{Notes:} Each cell reports the log-odds coefficient from a logistic regression where the row treatment is the base group and the column treatment is the comparison group. The upper triangle of the table is the inverse of the lower triangle (i.e., coefficient signs are flipped).} \\
\multicolumn{6}{l}{\footnotesize \sym{*} \(p<0.10\), \sym{**} \(p<0.05\), \sym{***} \(p<0.01\)}\\
\end{tabular}
\end{table}

\begin{table}[htbp]
\centering
% Define the symbol command for significance stars, as in your original file
\def\sym#1{\ifmmode^{#1}\else\(^{#1}\)\fi}

\caption{Pairwise Comparison of Treatment Effects on Choice Accuracy (Preference Type: LEX)}
\label{tab:pairwise_r2}
\begin{tabular}{@{}lccccc@{}}
\toprule
& \multicolumn{5}{c}{\textbf{Comparison Treatment}} \\
\cmidrule(l){2-6}
\textbf{Base Treatment} & SD-WEIGHT & SD-LEX & SD-DIRECT & SD-CHOICE & ACCURACY \\
\midrule
\textbf{SD-WEIGHT} & --- & 0.333\sym{***} & 0.266\sym{**} & 2.093\sym{***} & 2.315\sym{***} \\
& & & & & \\
\textbf{SD-LEX} & -0.333\sym{***} & --- & -0.068 & 1.760\sym{***} & 1.981\sym{***} \\
& & & & & \\
\textbf{SD-DIRECT} & -0.266\sym{**} & 0.068 & --- & 1.827\sym{***} & 2.049\sym{***} \\
& & & & & \\
\textbf{SD-CHOICE} & -2.093\sym{***} & -1.760\sym{***} & -1.827\sym{***} & --- & 0.221 \\
& & & & & \\
\textbf{ACCURACY} & -2.315\sym{***} & -1.981\sym{***} & -2.049\sym{***} & -0.221 & --- \\
\bottomrule
\multicolumn{6}{p{0.9\textwidth}}{\footnotesize \textit{Notes:} Each cell reports the log-odds coefficient from a logistic regression where the row treatment is the base group and the column treatment is the comparison group. The upper triangle of the table is the inverse of the lower triangle (i.e., coefficient signs are flipped).} \\
\multicolumn{6}{l}{\footnotesize \sym{*} \(p<0.10\), \sym{**} \(p<0.05\), \sym{***} \(p<0.01\)}\\
\end{tabular}
\end{table}

\begin{table}[htbp]
\centering
% Define the symbol command for significance stars, as in your original file
\def\sym#1{\ifmmode^{#1}\else\(^{#1}\)\fi}

\caption{Pairwise Comparison of Treatment Effects on Choice Accuracy (Preference Type: SEP)}
\label{tab:pairwise_r3}
\begin{tabular}{@{}lccccc@{}}
\toprule
& \multicolumn{5}{c}{\textbf{Comparison Treatment}} \\
\cmidrule(l){2-6}
\textbf{Base Treatment} & SD-WEIGHT & SD-LEX & SD-DIRECT & SD-CHOICE & ACCURACY \\
\midrule
\textbf{SD-WEIGHT} & --- & -0.264\sym{***} & 0.111 & 1.916\sym{***} & 0.768\sym{***} \\
& & & & & \\
\textbf{SD-LEX} & 0.264\sym{***} & --- & 0.375\sym{***} & 2.180\sym{***} & 1.032\sym{***} \\
& & & & & \\
\textbf{SD-DIRECT} & -0.111 & -0.375\sym{***} & --- & 1.805\sym{***} & 0.657\sym{***} \\
& & & & & \\
\textbf{SD-CHOICE} & -1.916\sym{***} & -2.180\sym{***} & -1.805\sym{***} & --- & -1.148\sym{***} \\
& & & & & \\
\textbf{ACCURACY} & -0.768\sym{***} & -1.032\sym{***} & -0.657\sym{***} & 1.148\sym{***} & --- \\
\bottomrule
\multicolumn{6}{p{0.9\textwidth}}{\footnotesize \textit{Notes:} Each cell reports the log-odds coefficient from a logistic regression where the row treatment is the base group and the column treatment is the comparison group. The upper triangle of the table is the inverse of the lower triangle (i.e., coefficient signs are flipped).} \\
\multicolumn{6}{l}{\footnotesize \sym{*} \(p<0.10\), \sym{**} \(p<0.05\), \sym{***} \(p<0.01\)}\\
\end{tabular}
\end{table}

\begin{table}[htbp]
\centering
% Define the symbol command for significance stars, as in your original file
\def\sym#1{\ifmmode^{#1}\else\(^{#1}\)\fi}

\caption{Pairwise Comparison of Treatment Effects on Choice Accuracy (Preference Type: COMP)}
\label{tab:pairwise_r4}
\begin{tabular}{@{}lccccc@{}}
\toprule
& \multicolumn{5}{c}{\textbf{Comparison Treatment}} \\
\cmidrule(l){2-6}
\textbf{Base Treatment} & SD-WEIGHT & SD-LEX & SD-DIRECT & SD-CHOICE & ACCURACY \\
\midrule
\textbf{SD-WEIGHT} & --- & -0.306\sym{***} & 0.347\sym{***} & 1.315\sym{***} & 1.124\sym{***} \\
& & & & & \\
\textbf{SD-LEX} & 0.306\sym{***} & --- & 0.652\sym{***} & 1.621\sym{***} & 1.430\sym{***} \\
& & & & & \\
\textbf{SD-DIRECT} & -0.347\sym{***} & -0.652\sym{***} & --- & 0.969\sym{***} & 0.778\sym{***} \\
& & & & & \\
\textbf{SD-CHOICE} & -1.315\sym{***} & -1.621\sym{***} & -0.969\sym{***} & --- & -0.191 \\
& & & & & \\
\textbf{ACCURACY} & -1.124\sym{***} & -1.430\sym{***} & -0.778\sym{***} & 0.191 & --- \\
\bottomrule
\multicolumn{6}{p{0.9\textwidth}}{\footnotesize \textit{Notes:} Each cell reports the log-odds coefficient from a logistic regression where the row treatment is the base group and the column treatment is the comparison group. The upper triangle of the table is the inverse of the lower triangle (i.e., coefficient signs are flipped).} \\
\multicolumn{6}{l}{\footnotesize \sym{*} \(p<0.10\), \sym{**} \(p<0.05\), \sym{***} \(p<0.01\)}\\
\end{tabular}
\end{table}

\begin{table}[htbp]
\centering
\caption{Exact Randomization-Inference $p$-values for Headline Comparisons}
\label{tab:ri}
\begin{tabular}{lccc}
\toprule
Comparison (vs.\ SD-DIRECT) & Choice accuracy & Efficiency loss & Justified envy \\
\midrule
ACCURACY  & 0.002 & 0.002 & 0.002 \\
SD-LEX    & 0.002 & 0.695 & 0.002 \\
SD-WEIGHT & 0.009 & 0.961 & 0.009 \\
SD-CHOICE & 0.002 & 0.002 & 0.002 \\
\bottomrule
\end{tabular}
\par\smallskip
\begin{minipage}{0.9\textwidth}
\footnotesize \textit{Notes:} Exact two-sided $p$-values from randomization inference. For each pairwise comparison, the treatment labels of the twelve independent matching groups (six per treatment) are permuted over all $\binom{12}{6}=924$ possible assignments, and the test statistic is the difference in matching-group mean outcomes. The smallest attainable two-sided $p$-value is $2/924\approx 0.002$. Outcomes are market-level choice accuracy, efficiency loss, and justified envy, averaged to the matching-group level; ACCURACY outcomes use simulated random-priority allocations as in Table~\ref{tab:accuracy-complexity}.
\end{minipage}
\end{table}

\begin{figure}[!ht]
    \centering
    \includegraphics[width=0.92\linewidth]{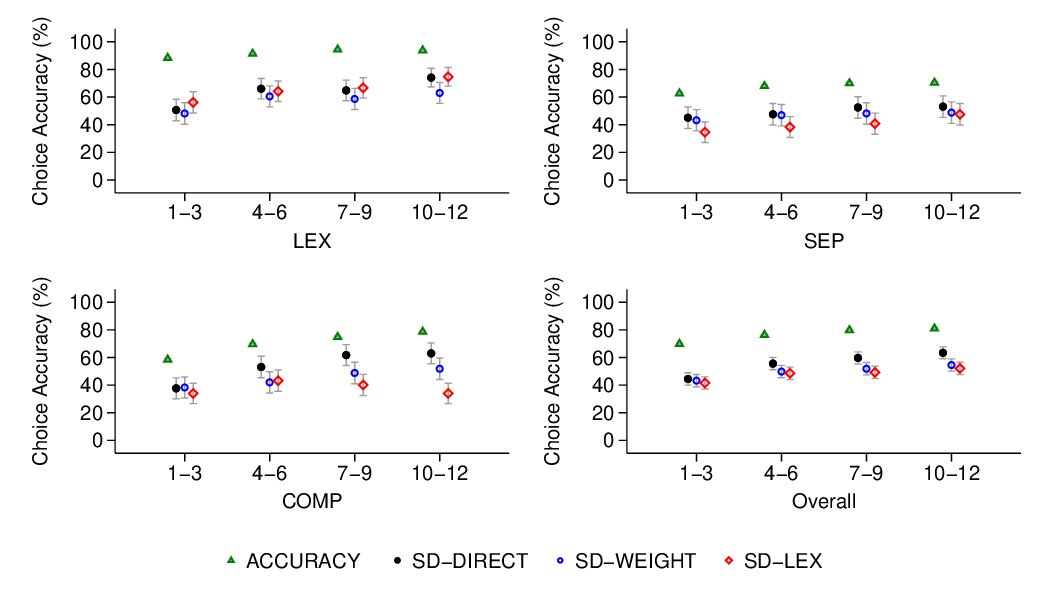}
    \caption{Choice Accuracy across Rounds (CI: 95\%)}
    \label{fig:rounds-acc}
            \par
    \footnotesize{Notes: X-axis reports rounds. Recall that the experiment design is such that each preference domain is realized exactly once in every three rounds.}
\end{figure}

\begin{figure}[!ht]
    \centering
    \includegraphics[width=0.95\linewidth]{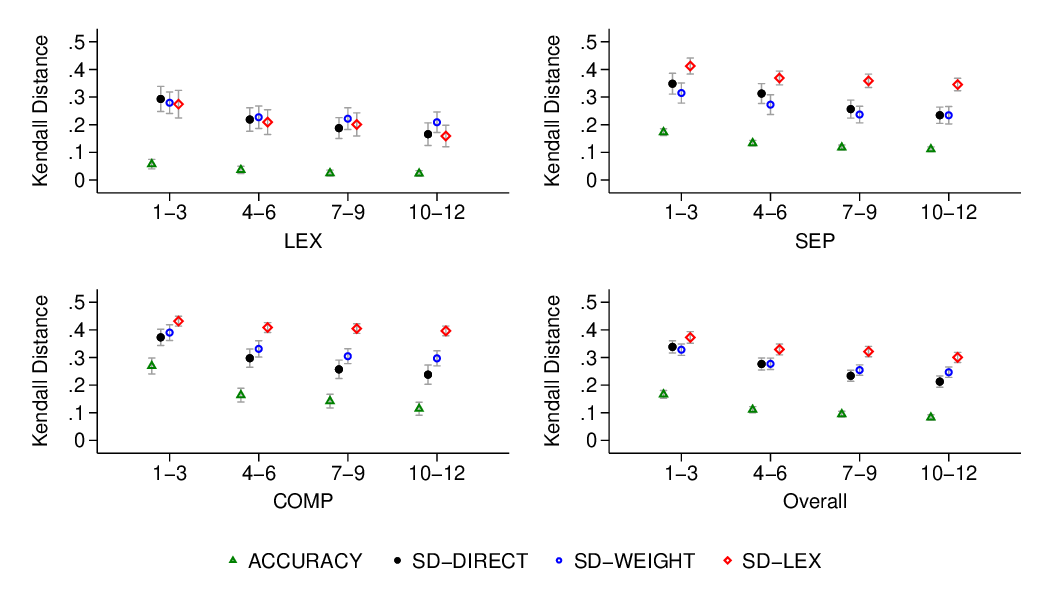}
    \caption{Kendall Distance across Rounds (CI: 95\%)}
    \label{fig:rounds-kd}
            \par
    \footnotesize{Notes: X-axis reports rounds. Each preference domain is realized exactly once in every three rounds.}
\end{figure}

\begin{figure}[!ht]
    \centering
    \includegraphics[width=1\linewidth]{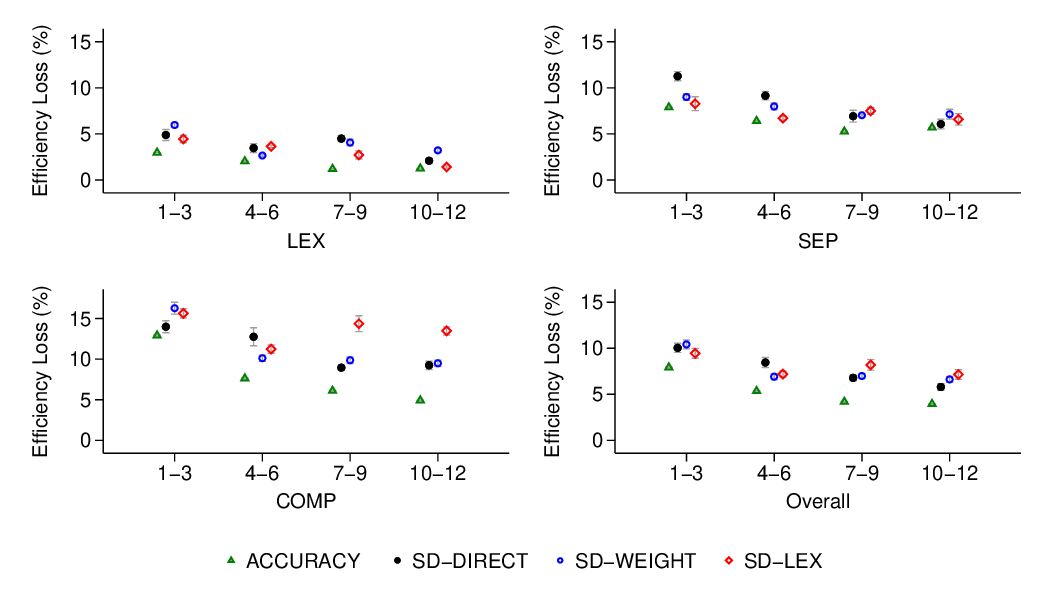}
    \caption{Efficiency Loss across Rounds (CI: 95\%)}
    \label{fig:rounds-el}
            \par
    \footnotesize{Notes: X-axis reports rounds. Each preference domain is realized exactly once in every three rounds.}
\end{figure}

\begin{figure}[!ht]
    \centering
    \includegraphics[width=1\linewidth]{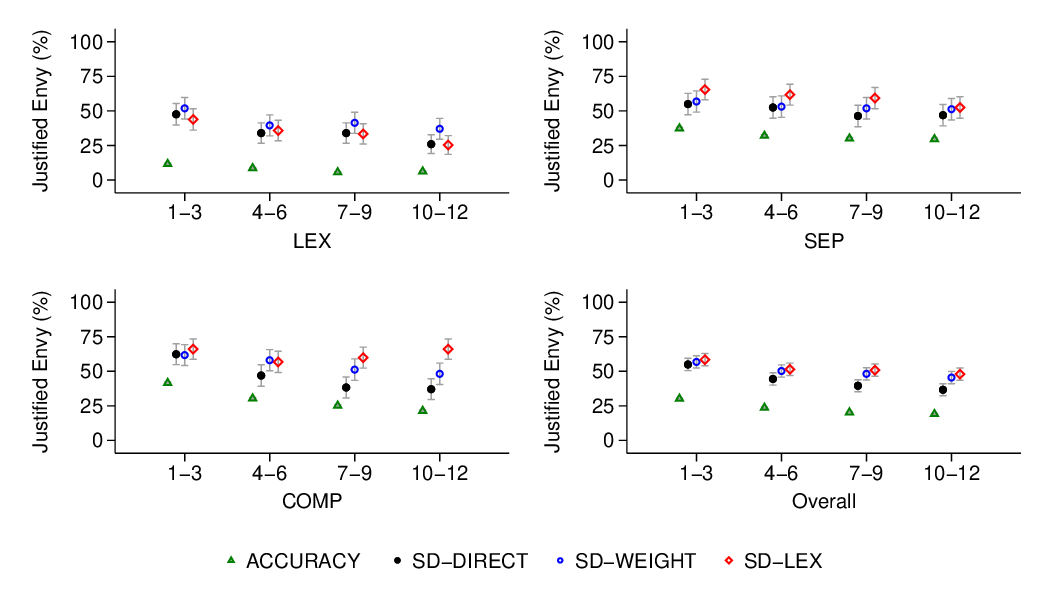}
    \caption{Justified Envy across Rounds (CI: 95\%)}
    \label{fig:rounds-je}
            \par
    \footnotesize{Notes: X-axis reports rounds. Each preference domain is realized exactly once in every three rounds.}
\end{figure}

\newpage 
\subsection{Computing Interface Expressiveness in Section~\ref{sec:expressiveness}}
\label{app:expressiveness}

For SD-LEX, we enumerate all $6^3 = 216$ combinations of attribute orderings; each yields a unique ranking of the 27 programs via the aggregator $\text{Points} = 100 \cdot \text{rank}(U) + 10 \cdot \text{rank}(F) + \text{rank}(T)$.

For SD-WEIGHT, we exhaustively enumerate the 216 attribute orderings combined with all integer weight triples in $[1,100]^3 = 10^6$, compute the implied ranking for each combination (completing ties by a fixed deterministic rule), and record the distinct rankings. This yields $58{,}176$ distinct rankings.

For each preference domain, we draw $10{,}000$ utility profiles from the parametric specifications in Section~\ref{sec:comppref}, compute the true induced ranking, and report the average minimum normalized Kendall distance between the true ranking and the closest expressible ranking under each interface. Weight combinations that generate ties in points do not induce a strict ranking (in the experiment, such ties were broken randomly). The floors are insensitive to how these combinations are treated in the enumeration: completing them with a fixed deterministic tie-breaking rule or excluding them altogether changes the average minimum distances by less than $0.02$. The floors are essentially unchanged when computed on the realized preference profiles of the experiment instead of simulated draws: averaged over all realized profiles, the minimum distances are $0.232$ (SEP) and $0.295$ (COMP) for SD-LEX, and $0.024$ (SEP) and $0.087$ (COMP) for SD-WEIGHT, with exact zeros on LEX. Code is available in the replication package.
\newpage
\section{Appendix: Indirect Message Spaces}
\subsection{Rank-order lists in College Admissions}
23 out of 31 provinces in China implement the structured rank-order list system, in which majors are effectively nested under colleges, as noted by \cite{huchinesecolleges}. These provinces include: Shanghai, Beijing, Tianjin, Hainan, Jiangsu, Fujian, Hubei, Hunan, Guangdong, Heilongjiang, Gansu, Jilin, Anhui, Jiangxi, Guangxi, Shanxi, Henan, Shaanxi, Ningxia, Sichuan, Yunnan, Tibet, and Xinjiang. For illustrative purposes, we include screenshots of the official college-major preference form from Fujian and Shanghai.

\begin{figure}[!ht]
    \centering
    \includegraphics[width=0.85\linewidth]{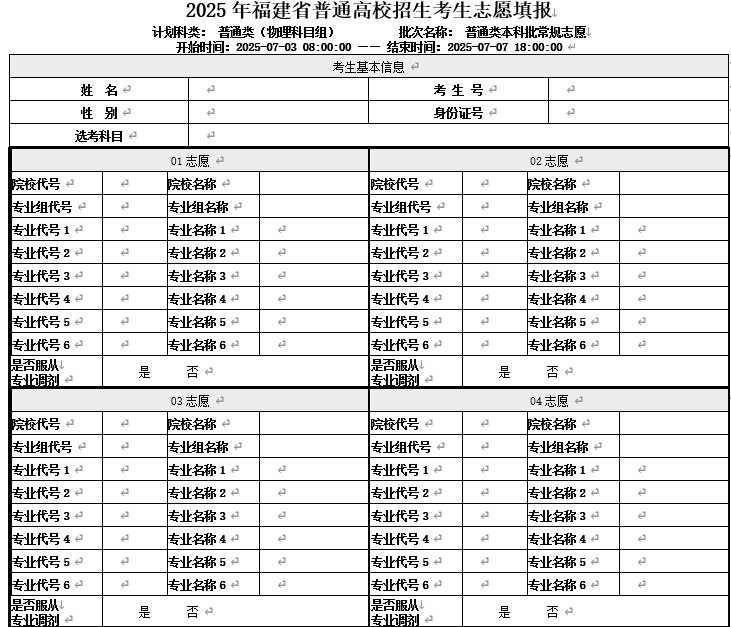}

    \floatfoot{\footnotesize Source: \url{https://www.eeafj.cn/gkptgkgsgg/20250626/14073.html}}
    %\caption{Caption}
    \label{fig:fujian_form}
\end{figure}
\newpage 

\begin{figure}[!ht]
    \centering
    \includegraphics[width=0.82\linewidth]{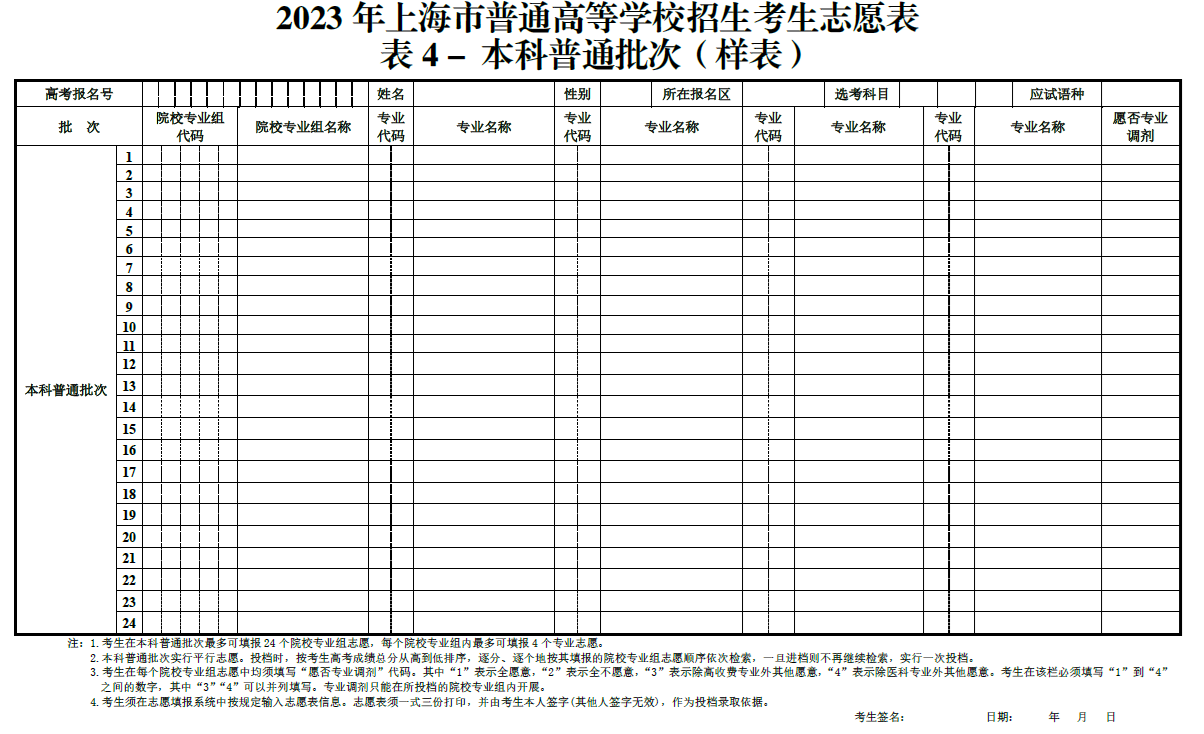}

        \vspace{-0.5cm} 
    
    \floatfoot{\footnotesize Source: \url{https://www.shmeea.edu.cn/page/08000/20230407/17353.html}}
    %\caption{Caption}
\end{figure}

\newpage

\subsection{Reserve Officer Training Corps (ROTC) Mechanism }

\cite{sonmez2013bidding}'s model of cadet-branch matching problem consists of
\begin{enumerate}
    \item a finite set of cadets $I = \{i_1, i_2, \dots, i_n\}$,
    \item a finite set of branches $B = \{b_1, b_2, \dots, b_m\}$,
    \item a vector of branch capacities $q = (q_b)_{b \in B}$,
    \item a set of ``terms" $T = \{t_1, \dots, t_k\}$,
    \item a list of cadet preferences $P = (P_i)_{i \in I}$ over $(B \times T) \cup \{\emptyset\}$, and
    \item a list of base priority rankings $\pi = (\pi_b)_{b \in B}$.
\end{enumerate}

The ROTC mechanism is not direct. Instead, each cadet submits a ranking of branches $\succ_i'$, and he can sign a branch-of-choice contract for any of his top three choices under $\succ_i'$. 

\newpage

\clearpage
\pagestyle{empty} 
\section{Appendix: Experimental Instructions (Translated)}
\noindent The instructions reproduced below are an English translation of the original Chinese instructions received by participants.
\clearpage
\includepdf[pages=-, pagecommand=\thispagestyle{empty}]{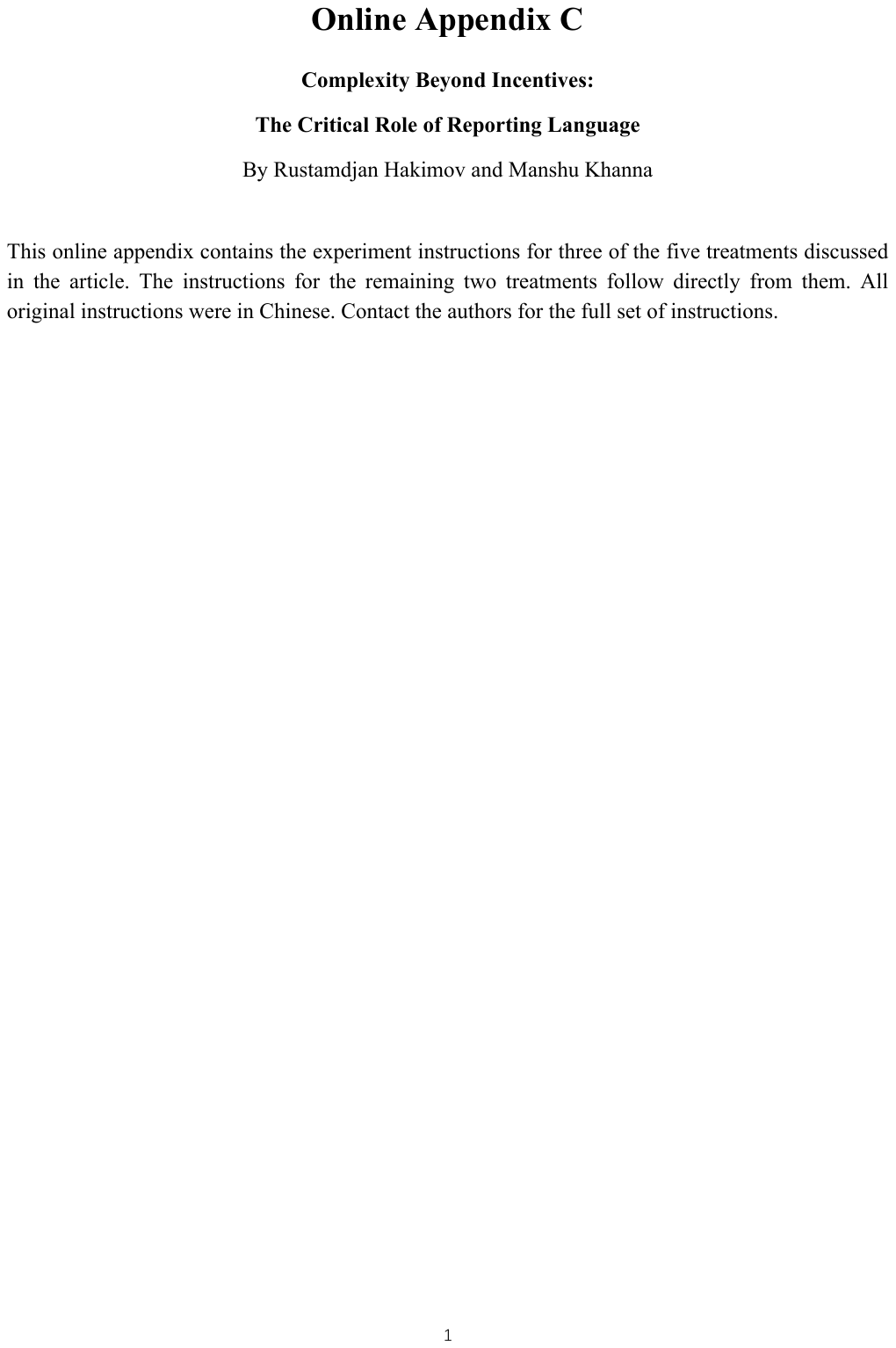}
\clearpage
\pagestyle{plain}

\end{document}